%% file: HM-arXiv-v2.tex
\documentclass[12pt,letter]{article}
\usepackage{amsfonts,latexsym,eucal,color,graphicx,epsfig,amssymb,graphicx,array,subfigure,cite}
\definecolor{red  }{rgb}{1,0,0}
\definecolor{blue }{rgb}{0,0,1}
\definecolor{green}{rgb}{0,1,0}
\begin{document}
\begin{titlepage}

\begin{flushright}
{\tt{WU-AP/286/08}}\\
\end{flushright}
\vspace{1cm}

%---------------------------------------------------------------------%
\begin{center}
{\huge Hidden symmetries, null geodesics, and photon capture in the Sen black hole}
\end{center}
\vspace{1cm}
%---------------------------------------------------------------------%

\begin{center}
\large{Kenta Hioki$^{1,\ast}$ and Umpei Miyamoto$^{2,\dagger}$}

%%--------   Address  ------------------
\vspace{.5cm}
{\small{\textit{$^{1}$Department of Physics, Waseda University, Okubo 3-4-1, Tokyo 169-8555, Japan \\
$^{2}$Racah Institute of Physics, Hebrew University, Givat Ram, Jerusalem 91904, Israel}}}
\\
%%----------------------------------------
\vspace*{1.0cm}
%%----------- Email  ----------------------
{\small
{\tt{$^{\ast}$hioki@gravity.phys.waseda.ac.jp\quad \quad $^{\dagger}$umpei@phys.huji.ac.il
}}
}
%%----------------------------------------
\end{center}

\vspace*{1.0cm}

\begin{abstract}
Important classes of null geodesics and hidden symmetries in the Sen black hole are investigated. First, we obtain the principal null geodesics and circular photon orbits. Then, an irreducible rank-two Killing tensor and a conformal Killing tensor are derived, which represent the hidden symmetries. Analyzing the properties of Killing tensors, we clarify why the Hamilton-Jacobi and wave equations are separable in this spacetime. We also investigate the gravitational capture of photons by the Sen black hole and compare the result with those by the various charged/rotating black holes and naked singularities in the Kerr-Newman family. For these black holes and naked singularities, we show the capture regions in a two dimensional impact parameter space (or equivalently the ``shadows'' observed at infinity) to form a variety of shapes such as the disk, circle, dot, arc, and their combinations.
\end{abstract}
\end{titlepage}

%---------------------------------------------------------------------%
%---------------------------------------------------------------------%
\section{Introduction}
\label{sec:intro}
%---------------------------------------------------------------------%
%---------------------------------------------------------------------%

In four dimensional effective theories of a higher dimensional gravity theory, the gravity couples to other fundamental fields such as the dilaton and gauge fields in various manners. The coupling manner affects the physical prediction especially in a strong gravitational regime, e.g., associated by black holes. The Sen black hole solution~\cite{Sen:1992ua} and Gibbons-Maeda black hole/brane solution~\cite{Gibbons:1987ps} in a low-energy limit of the heterotic string theory have been known as analytic solutions, which couple to the dilaton and gauge fields. It is important to know how the coupling manner affects the physical consequence in strong gravitational phenomena to construct a physically plausible theory of quantum gravity.

When one analyses the properties of black holes using geodesics, the separability and integrability of the geodesic equation are essential. Recently, much attention has been paid to the integrability of geodesic motion in four and higher dimensional black holes~\cite{Chen:2006ui, Frolov:2006pe, Vasudevan:2005js, Vasudevan:2004ca}. In general, it can happen that the number of conserved quantities of continuous isometries, which are projections of the geodesic momentum on Killing vector fields, are insufficient to integrate the geodesic motion. For example, the Killing vectors representing the stationarity and axisymmetry in the Kerr black hole are insufficient for the full integration of geodesic equations. One additional integral, however, is known to be given by a rank-two Killing tensor~\cite{Carter:1968rr}. Thus, the Killing tensor~\cite{Walker:1970un} as well as the Killing-Yano~\cite{Yano:1952} tensor play an essential role and have been obtained in various black hole solutions~\cite{Page:2006ka,Davis:2006hy,Kubiznak:2006kt,Chong:2004hw,Frolov:2003en,Ahmedov:2008pq}. Regarding the Sen black hole, however, it has not been known if such a Killing tensor exists, although the Hamilton-Jacobi and Klein-Gordon equations in this solution have been known to be separable~\cite{Blaga:2001wt} (both in the Einstein and string frames). In addition, important classes of null geodesics such as that of circular photon orbits or principal null geodesics have not been obtained for this solution. Therefore, it is interesting to obtain the Killing tensor, if any, and to know the properties of geodesics in the Sen black hole. In this paper, we are concerned to obtain them and to clarify the relation between the separability and symmetries.

As an application of the analysis of geodesics, we consider the capture and scattering of photons both by the Sen black hole and Gibbons-Maeda spacetimes, where the latter can be regarded as a nonrotating limit of the former. Then, the results are compared with those by the Kerr-Newman class spacetimes including both black hole and naked singular solutions. This problem is interesting~\cite{Gooding:2008tf, Zakharov:1994ts, Young:1976ca} because the capture region in an impact parameter space can be regarded as an apparent shape (shadow)~\cite{Zakharov:2005ek, Zakharov:2005sg, AdeVries:2000a} of the black hole to be observed at infinity. The shadows of black holes indeed can be utilized to obtain the physical information of strong gravity by a direct imaging~\cite{Takahashi:2004xh, Falcke:1999pj, Luminet:1979Aa}.

This paper is organized as follows. In Sec.~\ref{sec:solutions}, we review the several black hole solutions analyzed in this paper such as the Sen, Gibbon-Maeda, and Kerr-Newman class solutions. In Sec.~\ref{sec:geodesic}, we derive the geodesic equations in the Sen black hole and obtain the important classes of null geodesic. In Sec.~\ref{sec:Killing}, we obtain the irreducible rank-two Killing tensor and investigate their properties which are related to the separability. In Sec.~\ref{sec:capture}, we investigate the capture and scattering of photons by the Sen black hole and other charged/rotating black holes and naked singularities in the Kerr-Newman class. The final section is devoted to a summary. We use the geometric unit, in which $c=G=1$, throughout this paper.

%---------------------------------------------------------------------%
%---------------------------------------------------------------------%
\section{Sen black hole and other charged/rotating solutions}
\label{sec:solutions}
%---------------------------------------------------------------------%
%---------------------------------------------------------------------%

A low-energy effective action of the heterotic string theory is given by
\begin{eqnarray}
	I 
	=
	\frac{1}{16\pi} \int d^4x \sqrt{-g}
	\left[
		R - 2\left( \nabla \phi \right)^2
		- e^{-2\phi }F^2- \frac{1}{12} e^{-4\phi}H^2
	\right] \ .
\end{eqnarray}
Here, $\phi$ is a dilaton, and $F$ and $H$ are the field strength of an electromagnetic and an axion fields, respectively. The line element of the Sen black hole~\cite{Sen:1992ua} is given by
\begin{eqnarray}
	&&
	ds^2
	=
	-\left(
		\frac{\Delta -a^2\sin ^2 \theta}{\Sigma }
	\right) dt^2
	-
	\frac{4\mu r a \cosh ^2\beta \sin ^2 \theta}{\Delta} dtd\varphi
	+
	\frac{\Sigma}{\Delta } dr^2
	+
	\Sigma d\theta ^2
	+
	\frac{ \Lambda \sin^2\theta }{ \Sigma } d\varphi ^2 \ ,
\nonumber
\\
	&&
	A_{\mu}dx^{\mu} = \frac{\mu r \sinh 2\beta}{\sqrt{2}\Sigma}(dt-a\sin ^2\theta d\varphi ) \ ,
\;\;
	B_{t\varphi } = 2a^2 \sin ^2 \theta \frac{\mu r \sinh ^2 \beta}{\Sigma} \ ,
\nonumber
\\
&&
	\phi = -\frac{1}{2}\ln \frac{\Sigma}{r^2 +a^2\cos ^2 \theta} \ ,
\label{eq:metric}
\end{eqnarray}
where
\begin{eqnarray}
	&&
	\Delta
	:=
	r^2 - 2\mu r + a^2 \ ,
\;\;\;
	\Sigma
	:=
	r^2 + a^2 \cos^2 \theta +2\mu r\sinh ^2 \beta \ ,
	\nonumber
	\\
	&&
	\Lambda 
	:=
	\left[ r(r+2\mu \sinh \beta ^2)+a^2 \right] ^2 -\Delta a^2 \sin ^2\theta \ .
\end{eqnarray}
$A_{\mu}$ and $B_{\mu\nu}$ are the potentials of $ F $ and $ H $, respectively.
Constants $\mu$, $\beta$ and $a$ are related to the physical mass $M$, $U(1)$-charge $Q$, and angular momentum $J$ by
\begin{eqnarray}
&&
	M = \frac{\mu}{2}(1+\cosh 2\beta )\ ,
\;\;\;
	Q = \frac{\mu}{\sqrt{2}}\sinh ^2 2\beta\ ,
\;\;\;
	J = \frac{a\mu}{2}\left( 1+\cosh 2\beta \right)\ .
\end{eqnarray}
The regularity of the event horizon requires $ |J| \leq M^2 -Q^2/2 $.

We will compare later the properties of the Sen black hole with those of various charged/rotating solutions such as the Gibbon-Maeda (GM) and Kerr-Newman (KN) class solutions.
Thus, let us review these solutions and relations among them. The action of the Einstein-Maxwell-dilaton system is given by
\begin{eqnarray}
	I_{\mathrm{EMD}}
	=
	\frac{1}{16\pi}
	\int d^4x \sqrt{-g}
	\left[
		R - 2 \left( \nabla \phi \right)^2
		- e^{-2\alpha \phi } F^2
	\right] \ .
\label{eq:1}
\end{eqnarray}
Again, $\phi$ and $F_{\mu\nu}$ are the dilaton and $U(1)$ gauge field, respectively,
and $\alpha$ is their coupling constant. The GM and KN spacetimes are solutions in the system given by action (\ref{eq:1}).
The Sen solution (\ref{eq:metric}) without a rotation $(J=0)$ is reduced to the GM solution~\cite{Gibbons:1987ps} which is the solution of action (\ref{eq:1}) with $\alpha=1$. It can be shown that the GM solution with $ 0 \leq |Q| < \sqrt{2} M $ and $ |Q| =\sqrt{2}M $ represent a black hole and a naked singularity, respectively. The Sen solution without the charge $(Q=0)$ is reduced to the Kerr solution.
The KN spacetime is the solution of the Einstein-Maxwell theory, i.e., the system given by action~(\ref{eq:1}) with $\alpha=0$ and $\phi\equiv 0$. The line element of the KN solution is given by 
\begin{eqnarray}
&&
	ds^2
	=
	-\left(1-\frac{2Mr-Q^2}{\rho ^2}\right) dt^2
	-\frac{(2Mr-Q^2)2a\sin^2\theta}{\rho ^2}dtd\varphi
\nonumber
\\
&&
\hspace{8cm}
	+\frac{\rho ^2}{\Delta _{\mathrm{KN}}}dr^2+\rho ^2 d\theta ^2
	+\frac{A\sin^2\theta}{\rho ^2}d\varphi ^2 \ ,
\nonumber
\\
&&
	A_{\mu}dx^{\mu}
	=
	\frac{Qr}{\rho ^2 } ( dt - a\sin ^2 \theta d \varphi )\ ,
\end{eqnarray}
where
\begin{eqnarray}
&&
	\Delta _{\mathrm{KN}}
	:=
	r^2 - 2Mr + a^2 + Q^2 \ ,
\;\;
	\rho ^2
	:=
	r^2 + a^2 \cos^2 \theta \ ,
\nonumber
\\
&&
A :=
\left( r^2+a^2\right) ^2-\Delta _{\mathrm{KN}}a^2\sin^2\theta \ .
\end{eqnarray}
Again, the parameters $M$, $Q$, and $a$ are the mass, electric charge and specific angular momentum, respectively. The physical angular momentum of the spacetime is given by $ J := M a $.
The KN solution is reduced to the Reissner-Nordstr\"om (RN) solution and to Kerr solution in the neutral ($Q=0$) and nonrotating ($J=0$) limits, respectively.

\begin{center}
	\begin{figure}[t]
		\setlength{\tabcolsep}{ 50 pt }
		\begin{tabular}{ cc }
			\includegraphics[width=5cm]{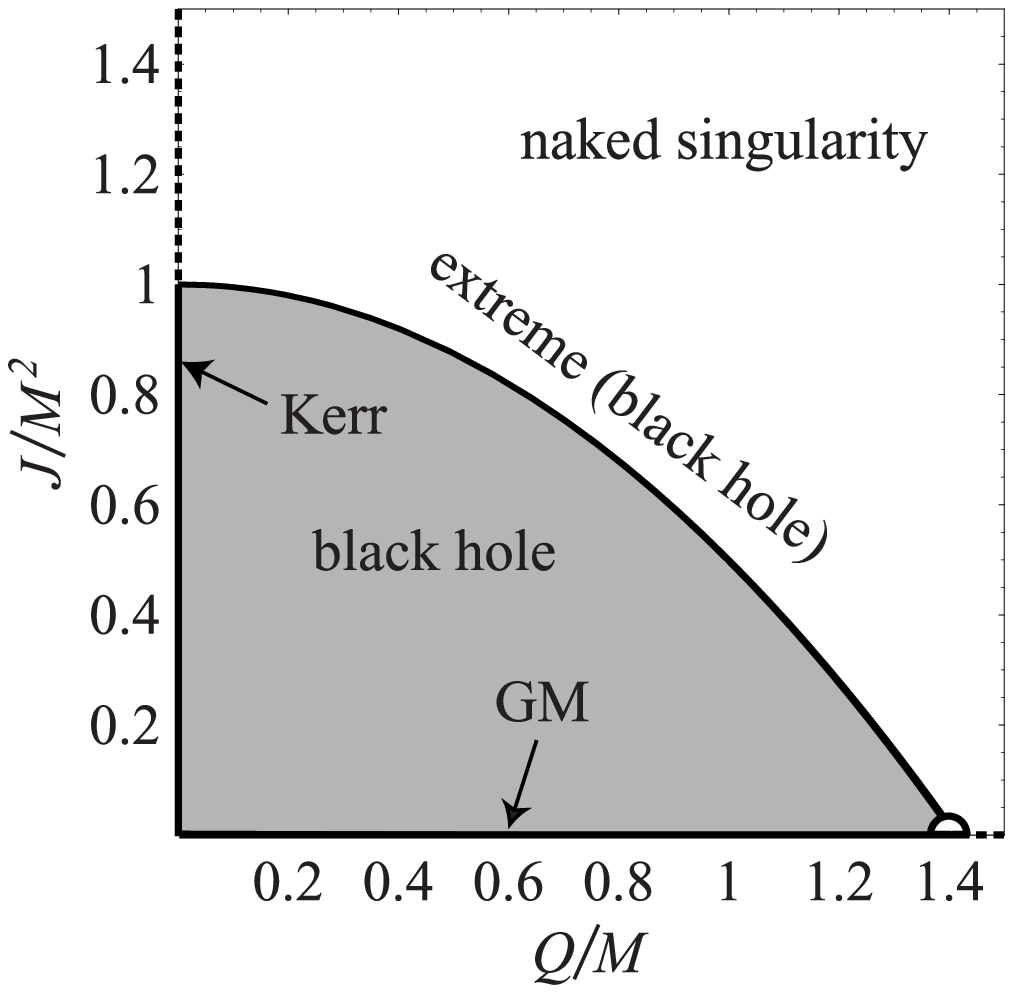} &
			\includegraphics[width=5cm]{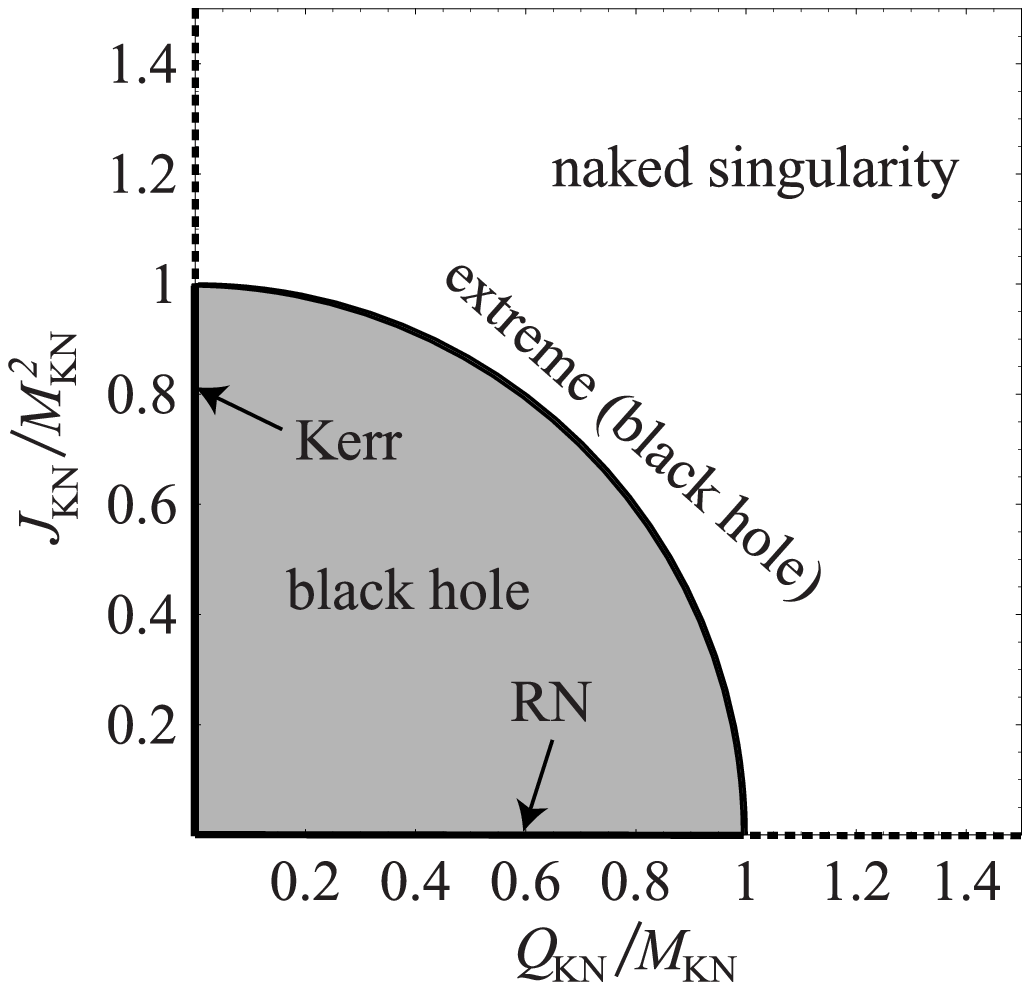} \\
			(a) Sen &
			(b) Kerr-Newman \\
		\end{tabular}
\caption{\textsf{\footnotesize{Part (a) shows the relation among the Sen, Gibbons-Maeda (GM), and Kerr solutions, and the boundaries separating the solutions into black hole solutions and naked singular solutions. The curve denoted by extreme corresponds to the extreme black hole, while the small semicircle at $Q =\sqrt{2}M$ corresponds to a naked singularity. Part (b) represents the relation among the Kerr-Newman (KN), Reissner-Nordstr\"om (RN), and Kerr solutions, and the boundaries separating the solutions into black hole solutions and naked singular solutions.
} } }				
		\label{fg:fig1}
	\end{figure}
\end{center}

In the rest of this paper, we denote the physical charges of KN, RN, and GM spacetimes by corresponding subscripts as $M_{\mathrm{KN}}$, $M_{\mathrm{RN}}$, and $M_{\mathrm{GM}}$. Quantities for the Sen solution are denoted without a subscript. The horizon regularity conditions in terms of charge and angular momentum for these solutions~\cite{Koga:1995bs} are summarized in Fig.~\ref{fg:fig1}.

%----------------------------------------------------------------------%
%----------------------------------------------------------------------%
%----------------------------------------------------------------------%
%----------------------------------------------------------------------%

%---------------------------------------------------------------------%
%---------------------------------------------------------------------%
\section{Null geodesics} 
\label{sec:geodesic}
%---------------------------------------------------------------------%
%---------------------------------------------------------------------%

%---------------------------------------------------------------------%
\subsection{Geodesic equations} 
\label{subsec:equation}
%---------------------------------------------------------------------%

The Hamilton-Jacobi equation for geodesic motion in the Sen black hole is written as
\begin{eqnarray}
	-\frac{{\partial S}}{{\partial \lambda}}
	=
	H
	=
	\frac{1}{2} g^{\mu \nu}\partial _\mu S\partial _\nu S  \ ,
\end{eqnarray}
where $S$ is the Hamilton principal function, $H$ is the Hamiltonian, and $\lambda$ is the affine parameter. We can separate variables by  
\begin{eqnarray}
	S 
	=
	\frac{1}{2} m^2 \lambda -Et+L_z+S_r(r)+S_{\theta}(\theta ) \ ,
\end{eqnarray}
where $S_r$ and $S_{\theta}$ are functions of $r$ and $\theta$, respectively. The constants $m$, $E$ and $L_z$ are the particle's mass, energy and angular momentum with respect to the rotation axis, respectively.
The $E$ and $L_z$ are conserved due to the existence of Killing vector fields $\partial _t$ and $\partial _\varphi$, respectively. The $r$- and $\theta$-dependent parts of the separated Hamilton-Jacobi equation~\cite{Blaga:2001wt} are
\begin{eqnarray}
	\Delta \left( \frac{{d S_r}}{{d r}} \right) ^2
	&=&
	\frac{1}{\Delta }
	\left\{
		\left[
			r(r+r_0)+a^2
		\right] E-aL_z
	\right\}^2
	-
	\left[
		K + m^2 r(r+r_0)
	\right] \ ,
\nonumber
\\
	\left(
		\frac{{d S_\theta}}{{d \theta}}
	\right)^2
	&=&
	K - \left(
			L_z \csc  \theta -Ea\sin \theta
		\right)^2
	-m^2 a^2 \cos ^2 \theta \ ,
\label{eq:separation}
\end{eqnarray}
where $K$ is a separation constant.

Differentiating $S$ with respect to $K$, $m^2$, $E$, and $L_z$, we have the following formal solutions in integral forms:
\begin{eqnarray}
	&&
	\sigma _r\int_{}^{r} \frac{dr}{\surd{R}}
	=
	\sigma _{\theta}\int_{}^{\theta} \frac{d\theta}{\surd{\Theta}}\ ,
	\nonumber	
	\\
	&&
	\lambda
	=
	\int_{}^{r} \frac{ r(r+r_0) }{ \surd{R} } dr
	+
	\int_{}^{\theta} \frac{ a^2 \cos^2 \theta }{ \surd{\Theta} } d\theta \ ,
	\nonumber
	\\
	&&
	t
	=
	\lambda E + \int_{}^{r} \frac{1}{\Delta \surd{R}}
	\Bigg[
		r
		\Big(
			 \left\{
				a^2 ( 2M + r_0 ) + r \left[ 2Mr + r_0 ( r + r_0 ) \right]
			  \right\} E
			- 2 M a L_z
		\Big)
	\Bigg] dr\ ,
	\nonumber
	\\
	&&
	\varphi
	=
	\int_{}^{r} \frac{a\left\{ \left[ r(r+r_0)+a^2\right] E-aL_z\right\} }{\Delta \surd{R}} dr
	+
	\int_{}^{\theta} \frac{ L_z\csc ^2\theta -aE }{ \surd{\Theta} }
	d\theta \ ,
\label{eq:int_geodecis}
\end{eqnarray}
where we have defined
\begin{eqnarray}
	&&
	R
	:=
	\left\{
		\left[
			r( r+r_0 ) + a^2
		\right] E - a L_z
	\right\}^2
	-
	\Delta \left[ (L_z-aE)^2+m^2 r(r+r_0)+\mathcal{Q}\right] \ ,
	\nonumber
	\\
	&&
	\Theta
	:=
	\mathcal{Q}-\left[ a^2\left( m^2-E^2\right) +L_z ^2 \csc ^2\theta \right] \cos ^2 \theta \ .
	\label{eq:RTheta}
\end{eqnarray}
Here, the sign functions, $\sigma _{r} =\pm$ and $\sigma _{\theta } =\pm$, are independent of each other and the sign changes at turning points of geodesic motion. The new parameter $r_0$ is defined by $r_0 := Q^2 /M$. We have also introduced a new constant of motion $\mathcal{Q}$ defined by $\mathcal{Q} := K-( L_z - Ea )^2$ and will use this hereafter for the constant $K$. This constant $\mathcal{Q}$ corresponds to the Carter constant in the Kerr solution.

It is also convenient to write Eq.~(\ref{eq:int_geodecis}) in the following first-order forms: 
\begin{eqnarray}
	&&
	\Sigma \frac{dr}{d \lambda}
	=
	\sigma _r\sqrt{R}\ ,
	\nonumber
	\\
	&&
	\Sigma \frac{d\theta}{d \lambda}
	=
	\sigma _{\theta} \sqrt{\Theta}\ ,
	\nonumber
	\\
	&&
	\Sigma \frac{d t}{d \lambda}
	=
	\frac{1}{\Delta}(\Lambda E-2MraL_z)\ ,
	\nonumber	
	\\
	&&
	\Sigma \frac{d\varphi}{d \lambda}
	=
	\frac{1}{\Delta} [2MarE+L_z\csc ^2\theta (\Sigma -2Mr)]\ .
	\label{eq:velocity}
\end{eqnarray}
Introducing new quantities by
\begin{eqnarray}
	\xi
	:=
	\frac{L_z}{E}\ ,
\;\;\;
	\eta
	:=
	\frac{\mathcal{Q}}{E^2}\ ,
\;\;\;
	\tilde R 
	:=
	\frac{R}{E^2}\ ,
\;\;
	\tilde \Theta
	:=
	\frac{\Theta}{E^2}\ ,
\;\;
	\tilde \lambda
	:=
	\lambda E\ ,
\end{eqnarray}
the energy $E$ can be eliminated from the system.
Then, the null geodesic ($m=0$) is parametrized only by ($\xi,\eta$)~\cite{Gyulchev:2006zg}. In terms of the parameters introduced above, Eq.~(\ref{eq:RTheta}) is rewritten as
\begin{eqnarray}
	\tilde R
	&=&
	\left[
		r( r + r_0 )
		+
		a^2-a\xi
	\right]^2
	-\Delta
	\left[
		\eta + ( a-\xi )^2
	\right]\ ,
\nonumber
\\
	\tilde \Theta
	&=&
	\eta + (a-\xi )^2 - ( a \sin \theta -\xi \csc \theta )^2
\nonumber
\\
	&=&
	\mathcal{I} - ( a \sin \theta -\xi \csc \theta )^2\ ,
\label{eq:potential}
\end{eqnarray}
where
\begin{eqnarray}
	\mathcal{I}(\xi,\eta) := \eta +(a-\xi )^2 \ . 
\end{eqnarray}
Since $\tilde R$ and $\tilde \Theta$ are non-negative from Eq.~(\ref{eq:int_geodecis}), the pair  of $(\xi,\eta)$ satisfies $ \mathcal{I}(\xi,\eta) \geq 0 $.

%---------------------------------------------------------------------%
\subsection{Principal null geodesics} 
\label{subsec:prin}
%---------------------------------------------------------------------%

We consider the null geodesics along the $\theta = \theta _0 = \mathrm{constant}$ plane. For such a photon motion, the set of parameters $(\xi,\eta)$ can be related to each other by
\begin{eqnarray}
	\tilde \Theta (\theta _0)
	=
	0\ ,
\;\;\;\;\;\;\;
	\partial _\theta \tilde \Theta \mid _{\theta _0}
	=
	0\ .
\label{eq:10}
\end{eqnarray}
Solving these equations while excluding trivial solutions, $\theta _0 =0$, $\theta _0 =\pi$, and $\theta _0 =\pi /2$, the set of parameters can be obtained as
\begin{eqnarray}
	\xi _\mathrm{prin}
	=
	a \sin^2 \! \theta _0 \ ,
\;\;\;\;\;\;\;
	\eta _\mathrm{prin}
	=
	- a^2 \cos^4 \! \theta _0 \ .
\label{eq:prin}
\end{eqnarray}
For later convenience, we note that instead of Eq.~(\ref{eq:10}), one can use the equation  $\mathcal{I}(\xi,\eta) = 0 $ to derive Eq.~(\ref{eq:prin}): since $\tilde \Theta$ is non-negative, $\mathcal{I} (\xi,\eta) = 0$ requires $ \tilde \Theta = 0 $, and so Eq.~(\ref{eq:prin}) is derived. It is also noted that we have $\tilde R = \Sigma ^2 > 0 $ if we substitute Eq.~(\ref{eq:prin}) into Eq.~(\ref{eq:potential}). This tells us that if the set $(\xi,\eta)$ satisfies $\mathcal{I}(\xi,\eta)=0$, the geodesic parametrized by $(\xi,\eta)$ has no turning point, and so must either approach the singularity or escape to the infinity.

The solution of Eq.~(\ref{eq:velocity}) for the set of parameters $(\xi_{\mathrm{prin}},\eta_{\mathrm{prin}})$ is the special null geodesic, i.e., the principal null geodesic, of which tangent vectors are given by 
\begin{eqnarray}
	l^{\mu}_{\pm} \partial _\mu
	:=
	\frac{1}{\Delta}
	\left\{
		\left[
			r(r+r_0)+a^2
		\right]
		\partial _t \pm \Delta \partial _r +a \partial _\varphi
	\right\} \ .
	\label{eq:l}
\end{eqnarray}
We can see that a congruence generated by $l^{\mu}_{\pm}$ is the nondegenerate principal null congruence since $l^{\mu} _{\pm}$ satisfies
\begin{eqnarray}
l_{\pm}^{\nu} l_{\pm}^{\sigma } l_{\pm[\epsilon }C_{\mu ]\nu \sigma [\delta }l_{\pm\gamma ]}=0 \ ,
\end{eqnarray}
where $C_{\mu \nu \sigma \delta }$ is the Weyl tensor.

Now let us see the properties of $l_{\pm}^\mu$, which are called the principal null vectors.
We consider the Kinnersley tetrad~\cite{1969:JMPK} $\left( l_{+}^{\mu}, n^{\mu}, m^\mu , \bar{m}^\mu \right)$, which are given by Eq.~(\ref{eq:l}) and
\begin{eqnarray}
&&
	n^\mu \partial _\mu
	:=
	\frac{1}{2\Sigma}
	\left\{
		\left[
			r(r+r_0)+a^2
		\right]	\partial _t
		- \Delta \partial _r + a\partial _\varphi
	\right\}\ ,
\nonumber
\\
&&
	m^\mu \partial _\mu
	:=
	\frac{1}
	{
	\sqrt{2}
	\left[
		\sqrt{ r(r+r_0)}+ia\cos \theta
	\right]
	}
	\left(
		ia\sin \theta \partial _t
		+
		\partial _\theta
		+
		i\csc \theta \partial _\varphi
	\right)\ .
\end{eqnarray}
These vectors satisfy the following orthonormal conditions:
\begin{eqnarray}
	l_{+}^\mu n_\mu
	=
	1 \ ,
\;\;\;\;\;\;
	m^\mu \bar{m}_\mu
	=
	-1 \ .
\end{eqnarray}
In addition, if we introduce the spin coefficients in the standard conventions of Newman-Penrose formalism~\cite{NP:1962},
\begin{eqnarray}
&&
	\kappa := -\left(\nabla _\mu l_{+\nu}\right) m^\nu l^\mu_{+}\ ,
\;\;\;\;\;\;\;
	\sigma := -\left(\nabla _\mu l_{+\nu}\right) m^\nu \bar m ^\mu\ ,
\nonumber
\\
&&
	\nu :=\left(\nabla _\mu l_{+\nu}\right) \bar m^\nu l^\mu _{+}\ ,
\;\;\;\;\;\;\;\;\;\;
	\lambda := \left(\nabla _\mu l_{+\nu}\right) \bar m^\nu m^\mu\ ,
\end{eqnarray}
one can show that these quantities vanish, and therefore that $l_{\pm}^\mu$ and $n^\mu$ generate shear-free geodesic congruences~\cite{Burinskii:1995hk} just like in the Kerr solution. Thus, we can say that, if the set of parameters $(\xi,\eta)$ satisfies $\mathcal{I}(\xi,\eta) =0$, the geodesic is the nondegenerate shear-free principal null geodesic\footnote{It is noted that the Goldberg-Sachs theorem~\cite{GS:1962} cannot be applied to determine the algebraical properties of the Sen black hole since this solution is not a vacuum solution.}.
These principal null geodesics will be used to investigate the symmetries of the spacetime in Sec.~\ref{sec:Killing}.

%---------------------------------------------------------------------%
\subsection{Circular photon orbits} 
\label{subsec:circular}
%---------------------------------------------------------------------%

In order to know the properties of radial motion, we analyze the function $\tilde R$ in Eq.~(\ref{eq:potential}). A circular photon orbit at constant radius $r=r_{\mathrm{circ}}$ satisfies the following conditions,
\begin{eqnarray}
	\tilde R(r_{\mathrm{circ}})
	=
	0\ ,
\;\;\;\;\;\;
	\partial _r \tilde R \mid  _{r_{\mathrm{circ}}}
	=
	0\ .
\label{eq:co}
\end{eqnarray}

First, we consider the nonrotating case $(J=0)$, where the Sen solution is reduced to the GM solution.
Solving Eq.~(\ref{eq:co}), the radius of the circular photon orbit is given by
\begin{eqnarray}
	r^{\mathrm{GM}}_{\mathrm{circ}}
	=
	\frac{1}{4}
	\left\{
		3(2M-r_0)
		+
		\left[
			(2M-r_0)(18M-r_0)
		\right] ^{1/2}
	\right\} \ .
\label{eq:GMun}
\end{eqnarray}
This circular photon orbit exists for $r_0 <2M$, i.e., $Q<\sqrt{2}M$, which means that the naked singularity in the GM solution has no circular photon orbit. Such a feature of GM solution is different from that of RN solution in that the naked singularity in the RN solution can have a circular photon orbit. One can also find that the circular photon orbits in the GM solution are unstable by calculating the second derivative of $\tilde{R}$. The existence and stability of circular photon orbits in the GM and RN solutions are summarized in Table.~\ref{table1}.

%---------------------------------------------------------------------%
%---------------------------------------------------------------------%
\begin{table}[b]
	\begin{center}
\caption{\textsf{\footnotesize{The regularity of spacetimes, stability of circular photon orbits, and shapes of the capture region (discussed in Sec.~\ref{sec:capture}) for the GM and RN solutions.
} } }				
	\vskip 5pt
\footnotesize{
	\setlength{\tabcolsep}{12pt}
		\begin{tabular}[c]{cc|cccccccc}
\hline
\hline
	\multicolumn{2}{c|}{Extremality $Q/M$}	&	\multicolumn{1}{c|}{0}	&	&	\multicolumn{1}{|c|}{1}	&
	&	\multicolumn{1}{|c|}{$3\sqrt{2}/4$}	&	&	\multicolumn{1}{|c|}{$\sqrt{2}$}	&	\\
\hline
	&	\multicolumn{1}{|c|}{Regularity}	&	\multicolumn{6}{c}{Black hole}	&	\multicolumn{2}{|c}{Naked singularity} \\
\cline{2-10}	
	GM	&	\multicolumn{1}{|c|}{Circular photon orbit}	&	\multicolumn{6}{c}{Unstable}	&	\multicolumn{2}{|c}{-----}	\\
\cline{2-10}
	&	\multicolumn{1}{|c|}{Capture region}	&	\multicolumn{6}{c}{Disk}	&	\multicolumn{2}{|c}{Dot} \\
\hline
	&	\multicolumn{1}{|c|}{Regularity} & \multicolumn{3}{c|}{Black hole}	&	\multicolumn{5}{c}{Naked singularity} \\
\cline{2-10}
	RN	&	\multicolumn{1}{|c|}{Outer circular photon orbit}	&	\multicolumn{4}{c}{Unstable}	&	\multicolumn{1}{|c|}{Marginal}	&	\multicolumn{3}{c}{-----} \\
\cline{2-10}
	&	\multicolumn{1}{|c|}{Capture region}	&	\multicolumn{3}{c|}{Disk}	&	\multicolumn{2}{c|}{Circle with dot}
	&	\multicolumn{3}{c}{Dot} \\
\hline
\hline
\end{tabular}
}
\label{table1}
\end{center}
\end{table}
%---------------------------------------------------------------------%
%---------------------------------------------------------------------%

For the nonrotating case ($J\not = 0$), we solve a quadratic equation for $\xi$, Eq.~(\ref{eq:co}), of which discriminant is given by $\mathcal{D}=(2r+r_0)^2(2r-2M+r_0)^2\Delta ^2$.
One set of solutions of Eq.~({\ref{eq:co}}) is 
\begin{eqnarray}
	\xi = \frac{r(r+r_0)+a^2}{a} \ ,
\;\;\;\;\;\;\;\;
	\eta = -\frac{r^2(r+r_0)^2}{a^2} \ .
\label{eq:20}
\end{eqnarray}
We can find, however, that this set of solutions is irrelevant as follows. One can show that this pair of parameters $(\xi,\eta)$ satisfies $\mathcal{I}(\xi,\eta)=0$. As we saw in Sec.~\ref{subsec:prin}, this implies $ \tilde R > 0 $, which is inconsistent with our condition Eq.~(\ref{eq:co}). Thus, the set of parameters for a circular photon orbit, we denote it by $(\xi_{\mathrm{circ}},\eta_{\mathrm{circ}})$, is given by the other solution of Eq.~({\ref{eq:co}}),
\begin{eqnarray}
&&
	\xi _{\mathrm{circ}}
	=
	\frac{1}{a(2r_{\mathrm{circ}}+r_0-2M)}
	\left[
		2M
		\left( r_{\mathrm{circ}}^2 -a^2\right)
		-
		\left( 2r_{\mathrm{circ}}+r_0\right) \Delta
	\right]\ , 
\nonumber
\\
&&
	\eta _{\mathrm{circ}}
	=
	\frac{r_{\mathrm{circ}}^2}{a^2(2r_{\mathrm{circ}}+r_0-2M)}
	\Big\{
		8a^2M(2r_{\mathrm{circ}}+r_0)
\nonumber
\\
&&
\hspace{6cm}
		-
		\left[
			(r_{\mathrm{circ}}+r_0-2M)(2r_{\mathrm{circ}}+r_0)-2Mr_{\mathrm{circ}}
		\right]^2
	\Big\}\ .
\label{eq:xietaco}
\end{eqnarray}
By calculating $\partial _r ^2 \tilde R \mid _{r_\mathrm{circ}}$ and paying attention to $\mathcal{D} \not = 0$, one can show that the unstable circular photon orbit $r=r_{\mathrm{circ}}$ exists for
\begin{eqnarray}
&&
	r_{\mathrm{circ}} < -\frac{r_0}{2} \ ,
\;\;\;\;\;
	M - M_0 - \frac{r_0}{2} < r_{\mathrm{circ}}  <  r_{-} \ ,
\;\;\;\;\;
	r_{+}  < r_{\mathrm{circ}} \ ,
\nonumber
\\
&&
	r_{\pm}
	=
	\frac{2M-r_0 \pm \left[ (2M-r_0)^2-4a^2 \right] ^{1/2} }{2} \ ,
%\;\;\;\;\;
\;\;
	M_0
	:=
	\left\{
		M \left[
			(2M-r_0)^2 -4a^2
		  \right] /4
	\right\}^{1/3} \ ,
\label{eq:region}
\end{eqnarray}
where $r_{+(-)}$ represents the location of the outer (inner) horizon. It should be noted that inequalities $ -r_0/2 \leq M-M_0-r_0/2 \leq r_{-} \leq r_{+}$ hold when the regular horizon condition $ |J| \leq M^2 -Q^2/2 $ is satisfied. These features of circular photon orbits will be used in Sec.~\ref{sec:capture}.

%---------------------------------------------------------------------%
%---------------------------------------------------------------------%
\section{Hidden symmetries} 
\label{sec:Killing}
%---------------------------------------------------------------------%
%---------------------------------------------------------------------%

%---------------------------------------------------------------------%
\subsection{Killing and conformal Killing tensors} 
\label{subsec:Killings}
%---------------------------------------------------------------------%

From Eq.~(\ref{eq:separation}) and the relation of $K=K^{\mu \nu} p_{\mu}p_{\nu}$ ($p_{\mu}$ is the four momentum of particle) we can find the following rank-two Killing tensor
\begin{eqnarray}
	K^{\mu \nu}
	=
	\delta ^{\mu}_{\theta} \delta ^{\nu}_{\theta}
	+
	\csc ^2\theta \delta ^{\mu}_{\varphi} \delta ^{\nu}_{\varphi}
	-
	2a \delta _\varphi ^{( \mu} \delta _t ^{{\nu )}}
	+
	a^2 \sin ^2 \theta \left( \delta ^{\mu}_{t} \delta ^{\nu}_{t}
	-
	\cot ^2 \theta g^{\mu \nu} \right) \ ,
\label{eq:Killing}
\end{eqnarray}
which satisfies $\nabla _{(\lambda } K_{\mu \nu )}=0$ and $K_{\mu\nu}=K_{(\mu\nu)}$. This Killing tensor $K^{\mu\nu}$ is independent of the linear combination of the symmetric outer product of Killing vectors $\partial _t$ and $\partial _\varphi $, and of the metric tensor. Therefore, $K^{\mu\nu}$ is irreducible. Using the Kinnersley tetrad, of which explicit expressions were given in Sec.~\ref{subsec:prin}, Eq.~(\ref{eq:Killing}) can be rewritten as
\begin{eqnarray}
	K_{\mu\nu}
	&=&
	2 \Sigma l_{+(\mu}n_{\nu )}
	+
	r(r+r_0)g_{\mu\nu} 
	\nonumber
	\\
	&=&
	2\left[ a^2 \cos ^2 \theta l_{+(\mu}n_{\nu )}
	+
	r(r+r_0) m_{(\mu}\bar{m}_{\nu )}\right]
	\ .
\label{eq:Killing2}
\end{eqnarray}

One can show that the two-rank tensor given by
\begin{eqnarray}
	Q_{\mu\nu}
	:=
	2 \Sigma l_{+(\mu}n_{\nu )}
\label{eq:CK}
\end{eqnarray}
is the conformal Killing tensor, satisfying $\nabla _{(\mu} Q_{\nu \lambda )} = g_{(\mu \nu}Q_{\lambda )}$ and $Q_{\mu\nu}=Q_{(\mu\nu)}$. Here, 
\begin{eqnarray}
	Q_\mu
	:=
	\frac{1}{6}
	\left(
		2 \nabla _\nu Q^\nu _{\ \mu}
		+
		\nabla _\mu Q^\nu _{\ \nu}
	\right)\ .
\end{eqnarray}
For the momentum of a null geodesic $p_{\mu}$, it is known that the quantity $Q_{\mu \nu} p^{\mu}p^{\nu}$ is conserved. The conformal Killing tensor is related to the Killing tensor by
\begin{eqnarray}
	K_{\mu\nu}
	=
	Q_{\mu\nu} - Q g_{\mu\nu}\ ,
\end{eqnarray}
where $ \nabla _\mu Q = Q_\mu $~\cite{Walker:1970un}.

We have seen that the Sen solution admits the irreducible Killing tensor and conformal Killing tensor. A comment is added here. In four dimensions, it is well known that any vacuum spacetime without an acceleration which is of type-D in the Petrov classification has a Killing tensor~\cite{Kubiznak:2007kh, Demianski:1980, Chandrasekhar:1985kt}. The Sen solution, however, is not algebraically special and is of type-I in the Petrov classification~\cite{Burinskii:1995hk}, and not vacuum solution. Therefore, the existence of a Killing tensor has not been trivial.

%---------------------------------------------------------------------%
\subsection{Separability structure} 
\label{subsec:sepa}
%---------------------------------------------------------------------%

We consider the origin of the separability of the Hamilton-Jacobi equation. Since a metric tensor $g_{\mu\nu}$ satisfies $\nabla _{(\lambda}g_{\mu\nu )}=0$, the metric tensor is a Killing tensor. We can show that this trivial Killing tensor $g_{\mu\nu}$ and the irreducible Killing tensor $K_{\mu\nu}$ mutually commute under the Schouten-Nijenhuis bracket~\cite{Benenti:1979},
\begin{eqnarray}
	[K_{\mu\nu}, g_{\mu\nu}]_\mathrm{S}
	:=
	K_{\sigma (\mu }\nabla ^\sigma g_{\nu \lambda )}-g_{\sigma (\mu }\nabla ^\sigma K_{\nu \lambda )}
	=
	0 \ .
\label{eq:SH}
\end{eqnarray}
The Killing tensors $(K_{\mu\nu},g_{\mu\nu})$ and Killing vectors $(\partial_t,\partial_\varphi)$ satisfy
\begin{eqnarray}
&&
	\pounds _{\partial _t} K_{\mu\nu}
	=
	0\ , 
\ \ 
	\pounds _{\partial _\varphi} K_{\mu\nu}
	=
	0\ ,
\nonumber
\\
&&
	\pounds _{\partial _t} g_{\mu\nu}
	=
	0 \ ,
\ \
	\pounds _{\partial _\varphi} g_{\mu\nu}
	=
	0 \ ,
\ \ 
	[\partial _t , \partial _\varphi] =0\ .
\label{eq:Lee}
\end{eqnarray}

Now, we show the existence of two independent simultaneous eigenvectors of the Killing tensors $K_{\mu\nu}$ and $g_{\mu\nu}$. One can show that the $\partial _r$ and $\partial _\theta$ are the eigenvectors of $K_{\mu\nu}$,
\begin{eqnarray}
&&
	K_{\mu\nu}(\partial _r)^\mu
	=
	- a^2 \cos ^2\theta (\partial _r)_\nu \ ,
\nonumber
\\
&&
	K_{\mu\nu} (\partial _\theta )^\mu
	=
	r( r+r_0 ) (\partial _\theta )_{\nu} \ .
\label{eq:Kl}
\end{eqnarray}
The Killing tensor $g_{\mu\nu}$ can be written as $g_{\mu\nu} = -2 \left( l_{+(\mu}n _{\nu )}- m_{(\mu} \bar{m}_{\nu)}\right)$. Therefore, $\partial _r$ and $\partial _\theta$ are also the eigenvectors of $g_{\mu\nu}$. Thus, $\partial _r$ and $\partial _\theta$ are the two independent simultaneous eigenvectors of the Killing tensors $K_{\mu\nu}$ and $g_{\mu\nu}$. In addition, it is obvious that the eigenvectors are mutually orthogonal, and the eigenvectors and the Killing vectors are mutually orthogonal. It is known that the above properties of the Killing tensors and Killing vectors guarantee the separation of variables in the Hamilton-Jacobi equation, which is called the separable structure~\cite{Houri:2007uq, Krtous:2006qy, Kalnins:1981ki}. As a consequence of the separability of the Hamilton-Jacobi equation, the additional conserved quantity $K$ is given, which mutually commutes with the Hamiltonian $H$ under the Poisson bracket. It means that we have four Poisson commuting functions $H$, $E$, $L_z$, and $K$. Thus, the system is completely integrable in the Liouville way as we saw in Eq.~(\ref{eq:int_geodecis}). Furthermore, one can show that the electromagnetic field $F$ satisfies the following relation:
\begin{eqnarray}
	K_{\mu (\nu} F_{ \;\;\lambda ) }^\mu
	=
	0\ .
\end{eqnarray}
This relation implies that the geodesic equation for a charged particle is also integrable~\cite{Hughston:1972qf}.

We now present a relation between the principal null vectors and Killing vector. We can write the covariant derivative of Killing vector $\partial  _t$ as
\begin{eqnarray}
	\nabla _{\mu} (\partial _t)_\nu
	=
	\partial _r \mathcal{F}	l_{+[\mu}n_{\nu ]}
	-
	\frac
	{
	2ia\cos \theta
	\left(
		1 -\mathcal{F}
	\right)
	}{\Sigma}
	m_{[\mu}\bar{m}_{\nu ]} \ ,
\end{eqnarray}
where $\mathcal{F} := - g_{\mu\nu} (\partial_t)^\mu (\partial_t)^\nu $. Thus, it is found that the principal null vector $l_{\pm}^\mu$ is the eigenvector of $\nabla _{\mu} (\partial _t)_\nu$,
\begin{eqnarray}
	\nabla _{\mu} (\partial _t)_\nu l_{\pm}^\mu
	=
	\frac{1}{2}\partial _r \mathcal{F} l_{\pm\nu} \ .
\label{eigen}
\end{eqnarray}
It has been known that such a relation as Eq.~(\ref{eigen}) plays an important role in the analysis of extended objects such as a string in the spacetime~\cite{Frolov:1995vp}. A similar result holds for the principal null vectors in the KN solution (e.g., see Ref.~\cite{Frolov:1995vp}).

%---------------------------------------------------------------------%
%---------------------------------------------------------------------%
\section{Gravitational capture of photons}
\label{sec:capture}
%---------------------------------------------------------------------%
%---------------------------------------------------------------------%

As an application of the analysis of null geodesics, we investigate the capture and scattering of photons by the Sen black hole and compare the result with those of other charged/rotating black holes and naked singularities described by the KN class solutions.

We take into account all photons launched from the infinity toward the black hole (or naked singularity) and examine which photons are captured or scattered away. In order to consider this problem, we have to introduce appropriate impact parameters. First, let us consider the tetrad components of photon's four momentum $\left( p^{(t)}, p^{(r)}, p^{(\theta)}, p^{(\varphi)}\right)$ with respect to a locally nonrotating reference frame~\cite{Bardeen:1972fi}. Using this tetrad, the set of impact parameters are defined by~\cite{Gooding:2008tf, Young:1976ca}
\begin{eqnarray}
&&
	b_x (\xi,\eta;i)
	:=
	\lim _{r \to \infty} 
	\frac{ r p^{(\varphi )} }{ p^{(r)} } = -\xi \csc i \ ,
\nonumber
\\
&&
	b_y (\xi,\eta;i)
	:=
	\lim _{ r \to \infty } \frac{ r p^{(\theta )} }{ p^{(r)} }
	=
	\sqrt{ \eta +a^2 \cos ^2 i -\xi ^2 \cot ^2 i} \ .
\label{eq:imapactparas}
\end{eqnarray}
Here, the photon is assumed to be parametrized by $(\xi,\eta)$, which are conserved quantities, and to be initially located at the infinity with a polar angle $ \theta = i $. Since the Sen black hole has the axisymmetry, the initial value of $\varphi$ can be fixed arbitrarily. Letting the parameters $(\xi,\eta)$ take all possible values [satisfying $\mathcal{I}(\xi,\eta)\geq 0$] with a fixed $i$, the capture region is given in the impact parameter space $(b_x, b_y)$ for the given $i$. It should be stressed that the capture region can be regarded also as the \textit{apparent shape} (or the \textit{shadow}) of the black hole/naked singularity, which is defined as the region in the impact parameter space not illuminated by the photon sources~\cite{AdeVries:2000a, Chandrasekhar:1985kt}.

%----------------------------------------------------------------------%
%----------------------------------------------------------------------%
\begin{center}
	\begin{figure}[t]
		\setlength{\tabcolsep}{ 10 pt }
		\begin{tabular}{ ccc }
			\includegraphics[width=3.5cm]{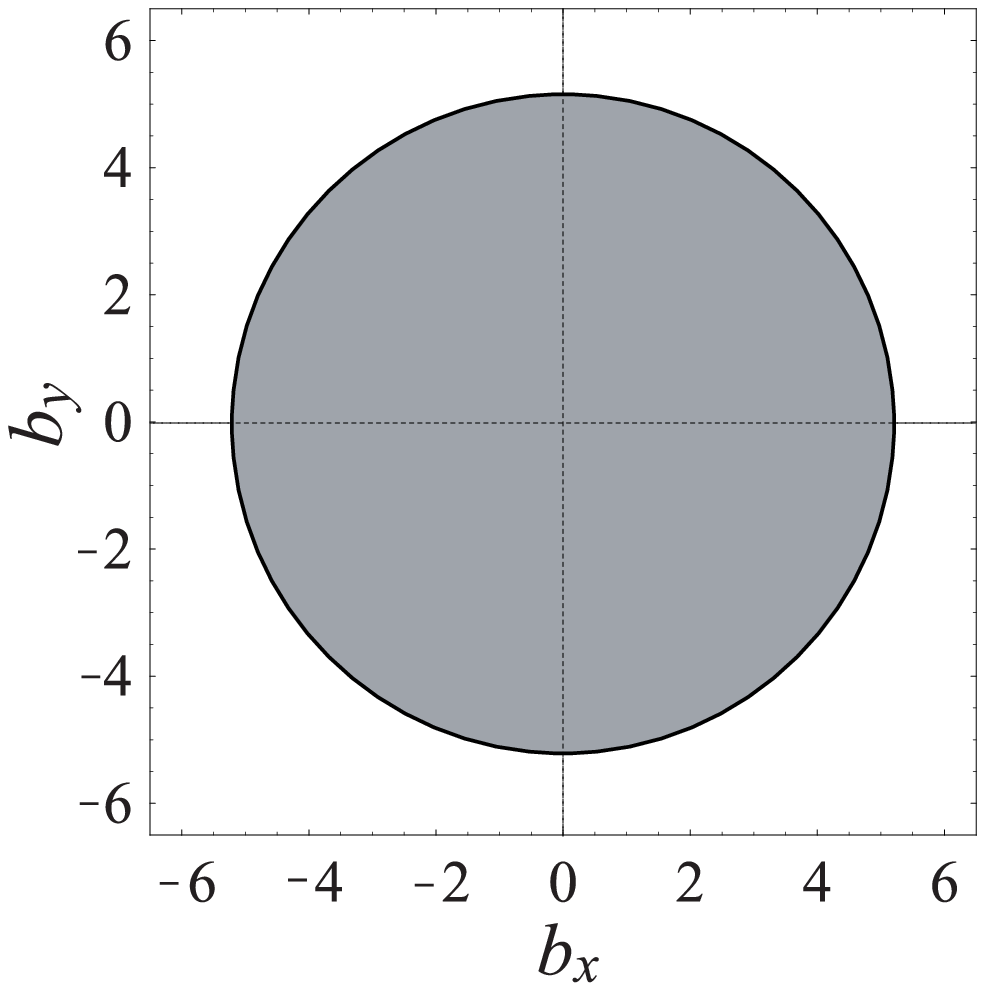} &
			\includegraphics[width=3.5cm]{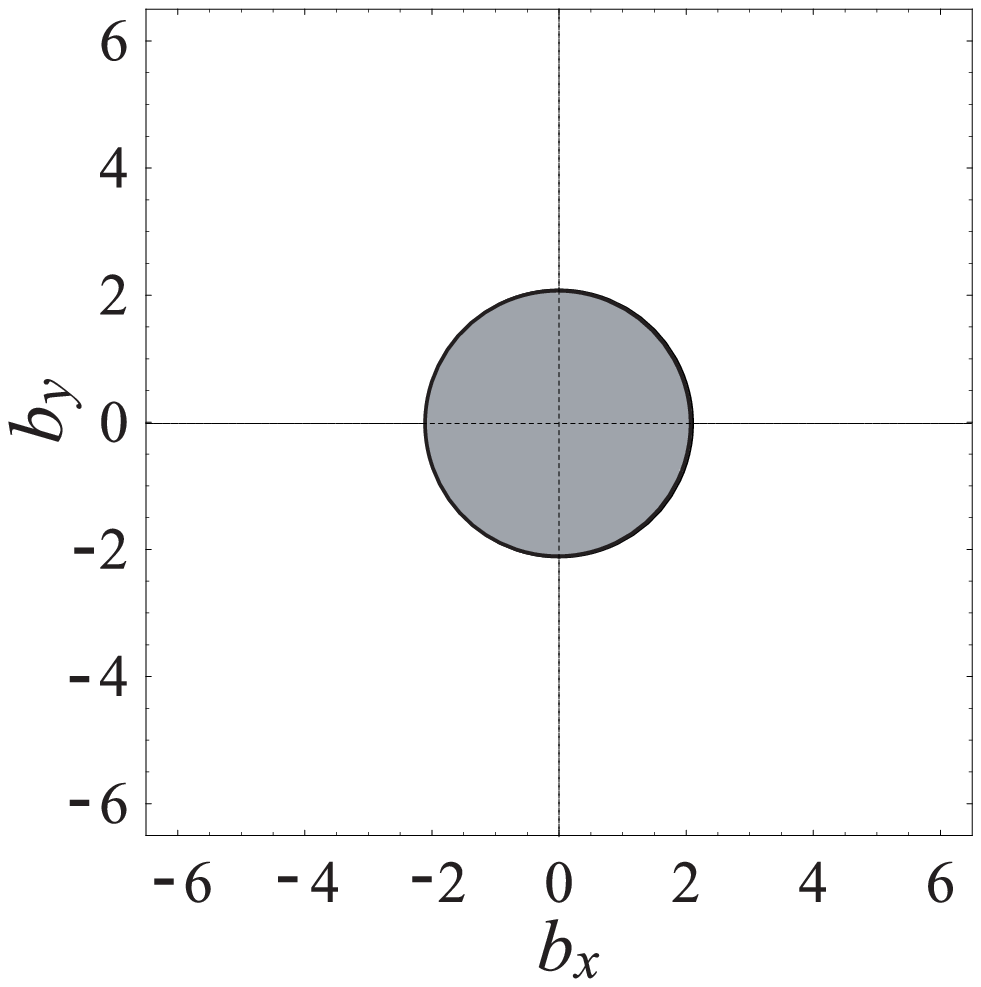} &
			\includegraphics[width=3.5cm]{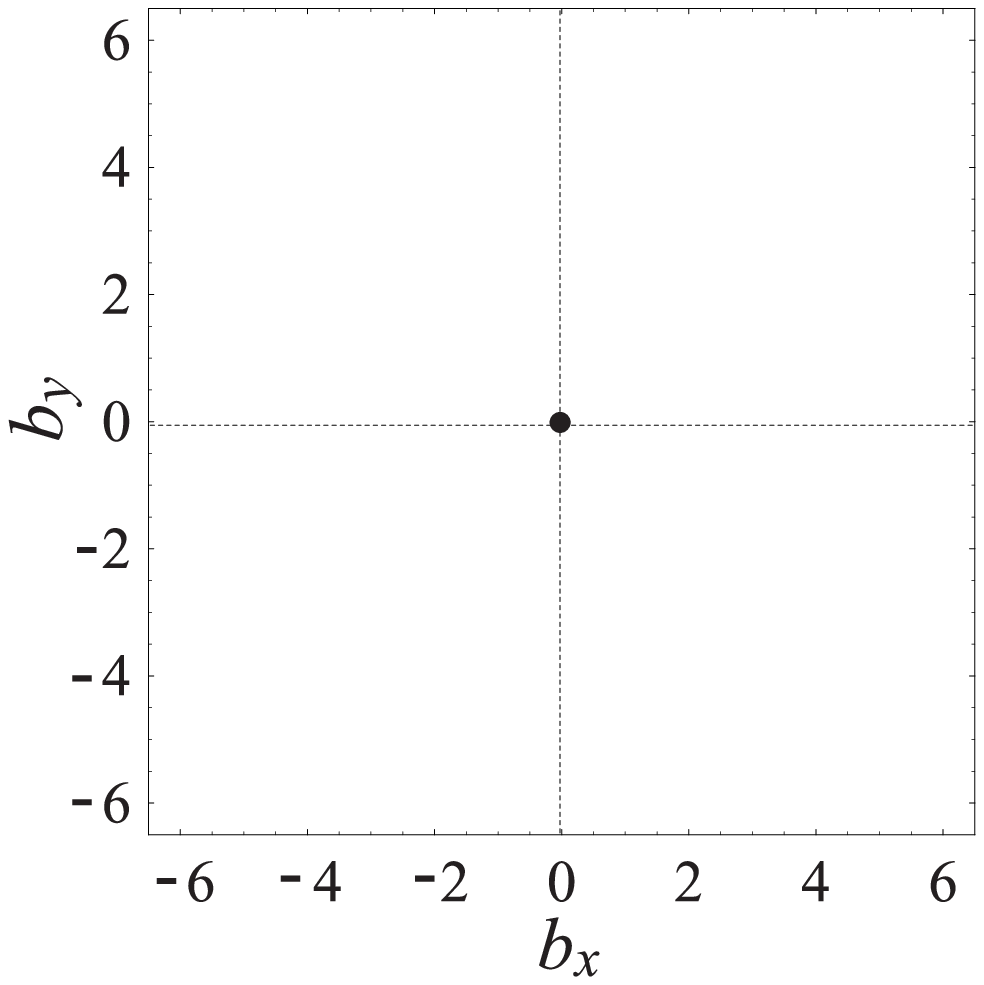} \\
			(a) $Q_\mathrm{GM}/M_\mathrm{GM}=0$ (BH)&
			(b) $Q_\mathrm{GM}/M_\mathrm{GM}=\sqrt{2}-10^{-4}$ (BH) &
			(c) $Q_\mathrm{GM}/M_\mathrm{GM}\geq \sqrt{2}$ (NS) \\
		\end{tabular}
\caption{\textsf{\footnotesize{Parts (a) and (b) show the capture regions in the impact parameter space $(b_x,b_y)$ for the GM black holes (BHs). Part (c) shows that for the GM naked singularity (NS). The unit of the coordinates is the mass $M_{\mathrm{GM}}$ in all cases. In the cases of black hole, (a) and (b), the photons are captured both by the circular photon orbit and event horizon to form the disks in the impact parameter space, while in the naked singularity case, (c), only the principal null geodesic is captured to form the dot. 
} } }				
		\label{fg:fig2}
%----------------------------------------------------------------------%
\vspace{.5cm}
%----------------------------------------------------------------------%
	\setlength{\tabcolsep}{ 45 pt }
		\begin{tabular}{ cc }
			\includegraphics[width=3.5cm]{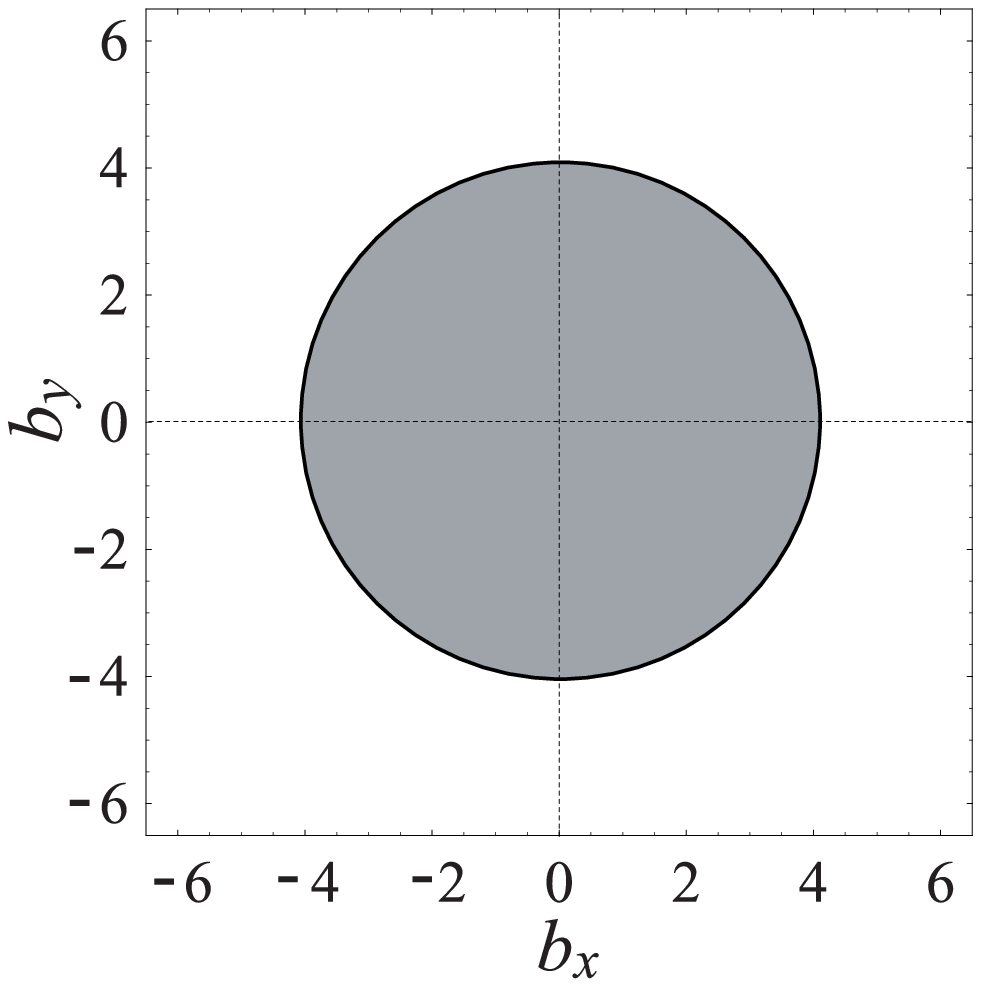} &
			\includegraphics[width=3.5cm]{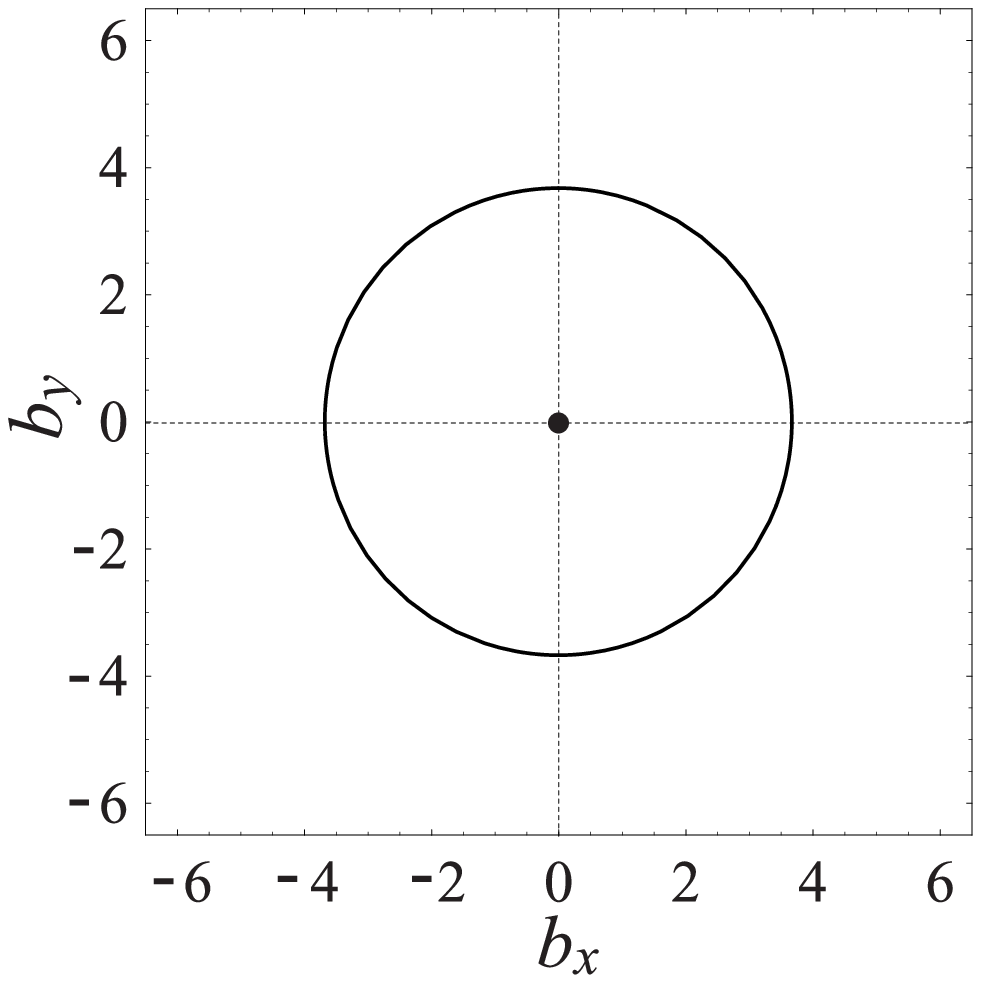} \\
			(a) $Q_\mathrm{GM}/M_\mathrm{GM}=3\sqrt{2}/4$ (BH) &
			(b) $Q_\mathrm{RN}/M_\mathrm{RN}=3\sqrt{2}/4$ (NS)\\
		\end{tabular}
\caption{\textsf{\footnotesize{Part (a) shows the capture region by the GM black hole. Part (b) shows the capture region by the RN naked singularity with the same extremality as the GM in (a). The unit of the coordinates is the mass $M_{\mathrm{GM}}$ and $M_{\mathrm{RN}}$, respectively. In the case of (a), the capture region forms the disk. While in the case of (b), the capture ``region'' is just the one-dimensional circle with the dot. This difference stems from the (non-)existence of the event horizon (see also Table.~\ref{table1}).
} } }				
		\label{fg:fig3}
	\end{figure}
\end{center}
%----------------------------------------------------------------------%
%----------------------------------------------------------------------%

%---------------------------------------------------------------------%
\subsection{Nonrotating case} 
\label{subsec:nonrotating}
%---------------------------------------------------------------------%

We first consider the nonrotating case ($J=0$), in which the Sen solution is reduced to the GM solution.
Because of the spherical symmetry, we can set the initial value of the polar angle to $i = \pi /2$ without loss of generality. Characteristic examples of the capture regions for the GM black hole and naked singularities are presented in Fig.~\ref{fg:fig2}. In addition, we compare the capture regions between the GM and KN solutions of the same extremality in Fig.~\ref{fg:fig3}. In the nonrotating cases containing the KN solutions, we find that the shapes of the capture region can be classified into three types: a disk, a dot, and a circle with a dot. The type of the capture region for a given charge in the cases of the GM and RN solutions are summarized in Table~\ref{table1}. The existence of the unstable circular photon orbits, principal null congruence, and event horizon play important roles to determine the capture region. In the following, we will explain how the shape of the capture region is determined by these characteristic null rays. The readers who are interested only in the result can skip to Sec.~\ref{subsec:rotating}. 

\begin{center}
	\begin{figure}[t]
		\setlength{\tabcolsep}{ 12 pt }
		\begin{tabular}{ ccc }
			\includegraphics[width=3.5cm]{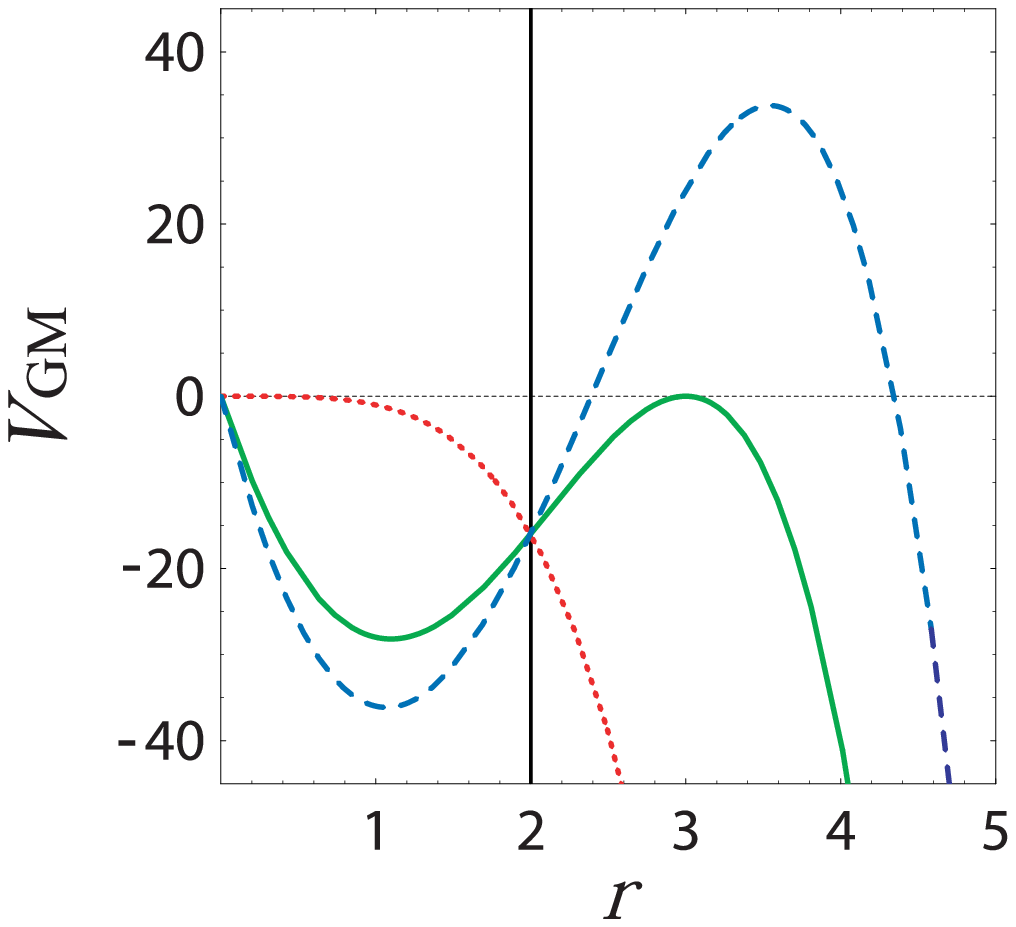} &
			\includegraphics[width=3.5cm]{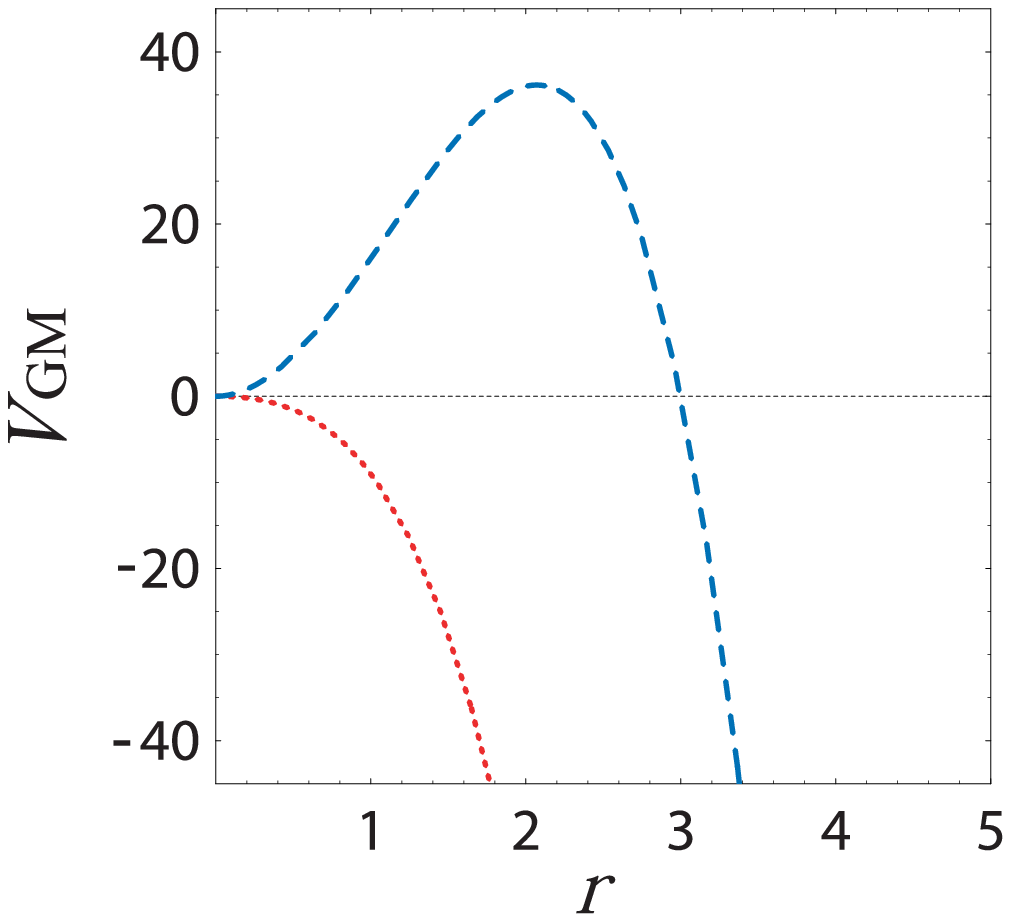} &
			\includegraphics[width=3.5cm]{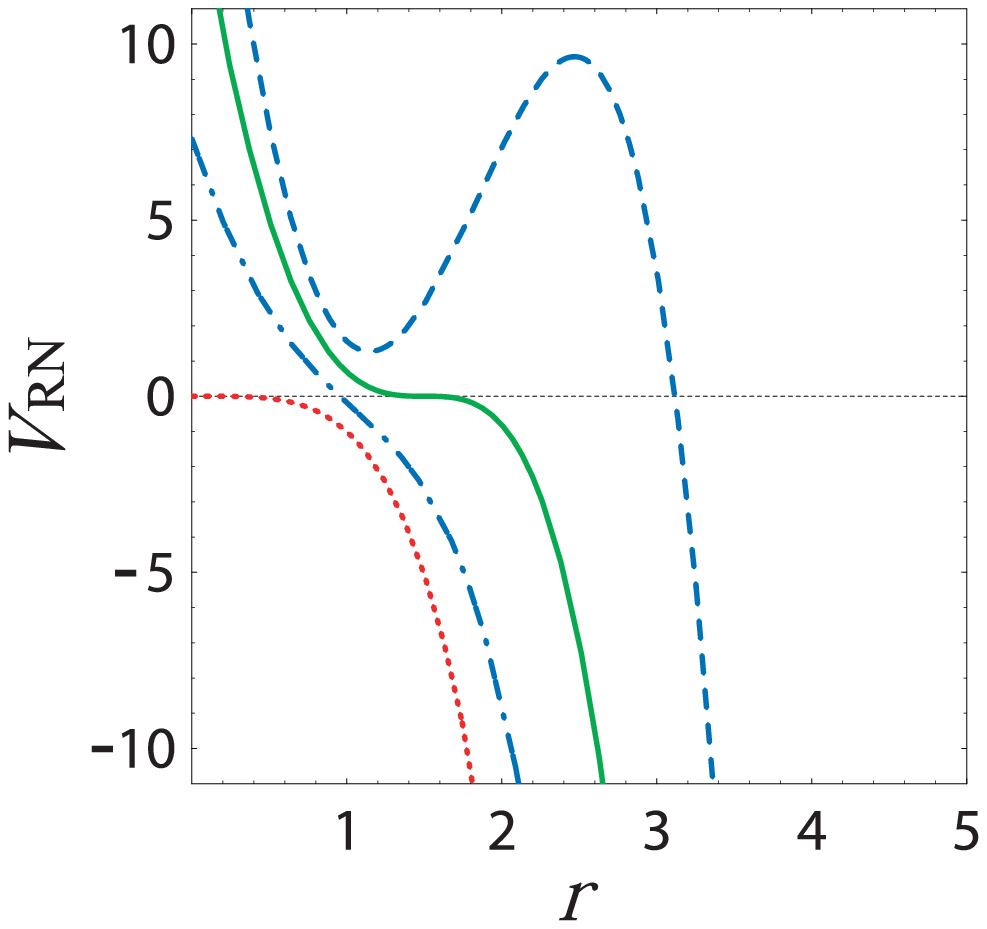} \\
			(a) $Q_\mathrm{GM}/M_\mathrm{GM}=0$ (BH) &
			(b) $Q_\mathrm{GM}/M_\mathrm{GM}=\sqrt{2}$ (NS) &
			(c)	$Q_\mathrm{RN}/M_\mathrm{RN}=3\sqrt{2}/4$ (NS)\\
		\end{tabular}
\caption{\textsf{\footnotesize{Potentials for the GM black hole, (a), those for the GM naked singularity, (b), and those for the RN naked singularity, (c). See Figs.~\ref{fg:fig2}(a), \ref{fg:fig2}(c), and \ref{fg:fig3}(b) for the corresponding shadows, respectively.
(a) The solid curve represents the potential for the unstable circular photon orbit, which forms the boundary of the shadow. The dotted curve represents the potential for the null geodesic forming the inner region of shadow. The dashed curve represents the potential for the null geodesic scattered away, which has a turning point outside the event horizon (normalized to being located at $r=2$).
(b) The dotted curve represents the potential for the principal null geodesic, which forms the dot in the shadow. While the potential for the null geodesic which is scattered away is plotted by the dashed curve.
(c) The dotted (solid) curve represents the potential for the principal null geodesic (marginally stable circular photon orbit), which forms the dot (circle) in the shadow.
The dashed (dot-dashed) curve represents the potential for the null geodesic corresponding to the outer (inner) region of the circle in the shadow.
} } }				
\label{fg:fig4}			
	\end{figure}
\end{center}

We can rewrite geodesic Eq.~(\ref{eq:velocity}) in the radial direction in a potential form,
\begin{eqnarray}
	\Sigma^2 \left( \frac{dr}{d\tilde \lambda} \right)^2 + V(r)=0\ ,
	\;\;\;\;\;\;
	V(r) := - \tilde{R}\ .
\label{eq:pote-form}
\end{eqnarray}
It is straightforward to write down the null geodesic equation in the radial direction also for the RN solution in the form of Eq.~(\ref{eq:pote-form}). The explicit expressions of potential for the GM and RN solutions are 
\begin{eqnarray}
&&
	V_\mathrm{GM}(r)
	=
	- r
	\left[
		r(r+r_0)^2 - ( r-2M_\mathrm{GM}+r_0 )
		\left(
			\eta +\xi ^2
		\right)
	\right]\ ,
\nonumber
\\
&&
	V_\mathrm{RN}(r)
	=
	- r^4
	+
	\left[
		Q_\mathrm{RN}^2+r(r-2M_\mathrm{RN})
	\right]
	\left( 
		\eta +\xi ^2
	\right)\ .
\end{eqnarray}
When we fix the spacetime parameters (mass and charge in this case), the potential depends on the set of parameters of particle, $(\xi,\eta)$.
We show some schematic pictures of potential in Fig.~\ref{fg:fig4}, in each of which the potentials for several sets of $(\xi,\eta)$ are plotted: Fig.~\ref{fg:fig4}(a) corresponds to the potentials in the GM black hole of which the capture region is given in Fig.~\ref{fg:fig2}(a); Fig.~\ref{fg:fig4}(b) corresponds to the potentials in the GM naked singularity of which the capture region is given in Fig.~\ref{fg:fig2}(c); and Fig.~\ref{fg:fig4}(c) corresponds to the potentials in the RN naked singularity of which the capture region is given in Fig.~\ref{fg:fig3}(b).

In the case where both the event horizon and circular photon orbit exist [e.g., see Figs.~\ref{fg:fig2}(a) and \ref{fg:fig4}(a)], the boundary of shadow is formed by the null geodesics which twine around the unstable circular photon orbit. On the other hand, the inner region of shadow is formed by the null geodesics which are captured and fall into the event horizon. In Fig.~\ref{fg:fig4}(a), the potentials for these null geodesics and that for the null geodesic which is scattered away are plotted. Now, we briefly describe how to draw the shadow using the results obtained in the previous sections. We obtained the radius of unstable circular photon orbit $r_{\mathrm{circ}}$ in Eq.~(\ref{eq:GMun}), therefore we can know the corresponding conserved quantities $(\xi_{\mathrm{circ}},\eta_{\mathrm{circ}})$ from Eq.~(\ref{eq:co}). Then, using the relation between impact parameters $(b_x, b_y)$ and ($\xi,\eta$) in Eq.~(\ref{eq:imapactparas}), we can plot the boundary of shadow in the impact parameter space $(b_x, b_y)$. Taking into account the null geodesics captured by the event horizons, we finally obtain the shadow as the disk as Fig.~\ref{fg:fig2}(a).

%----------------------------------------------------------------------%
%----------------------------------------------------------------------%
%----------------------------------------------------------------------%
%----------------------------------------------------------------------%

When neither the event horizon nor unstable circular photon orbit exist [e.g., see Figs.~\ref{fg:fig2}(c) and \ref{fg:fig4}(b)], the null geodesic with any set of conserved quantities $(\xi,\eta)$ has a turning point and is not captured except the principal null geodesic, which inevitably plunges into the singularity. Since we know $(\xi _\mathrm{prin},\eta _\mathrm{prin})=(0,0)$ for the nonrotating case from Eq.~(\ref{eq:prin}), we have $(b_x, b_y)=(0,0)$ for the principal null geodesic from Eq.~(\ref{eq:imapactparas}). Thus, the capture region in the GM naked singularity is the dot as Fig.~\ref{fg:fig2}(c). From another point of view, this means that the capture cross section of photons by the GM naked singularity vanishes.

When the unstable (or a marginally stable) circular photon orbit exists but the event horizon does not [e.g., see Figs.~\ref{fg:fig3}(b) and \ref{fg:fig4}(c)], the captured null geodesics are only the unstable circular photon orbit and ingoing principal null geodesic. Thus, the capture region is the circle with the dot as Fig.~\ref{fg:fig3}. The circle is formed by the unstable circular photon orbits, while the null geodesics corresponding to the inner or outer region of circle has the turning point to go back to the infinity except for the principal null geodesic, which forms the dot in the center. We find such situation that a circular photon orbit exists but the event horizon does not occur in the GM solution, while the naked singular RN solution (see Table.~\ref{table1}) allows this situation. It is noted that the capture region of the RN spacetimes in black hole cases was obtained in~\cite{Zakharov:1994ts, Zakharov:2005sg}.

%----------------------------------------------------------------------%
%----------------------------------------------------------------------%
\begin{center}
	\begin{figure}[t]
		\setlength{\tabcolsep}{ 25 pt }
		\begin{tabular}{ ccc }
			\includegraphics[width=3.5cm]{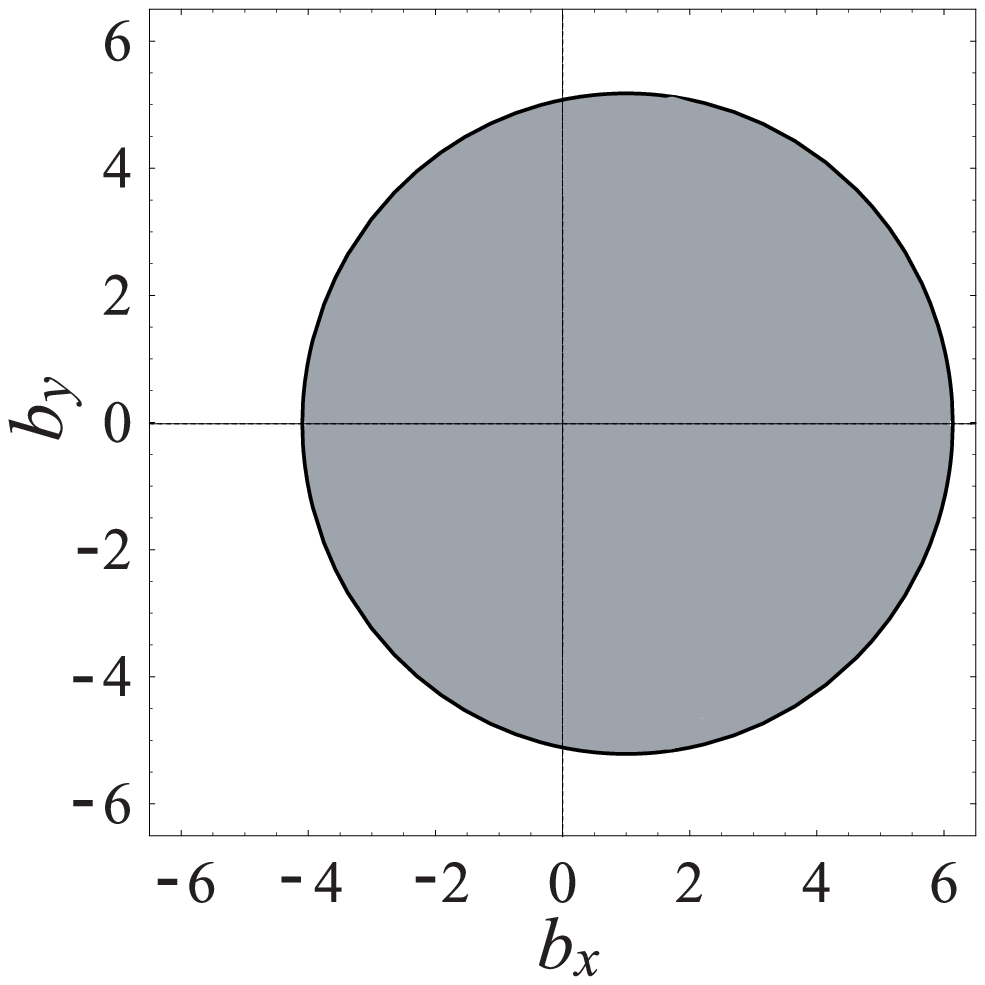} &
			\includegraphics[width=3.5cm]{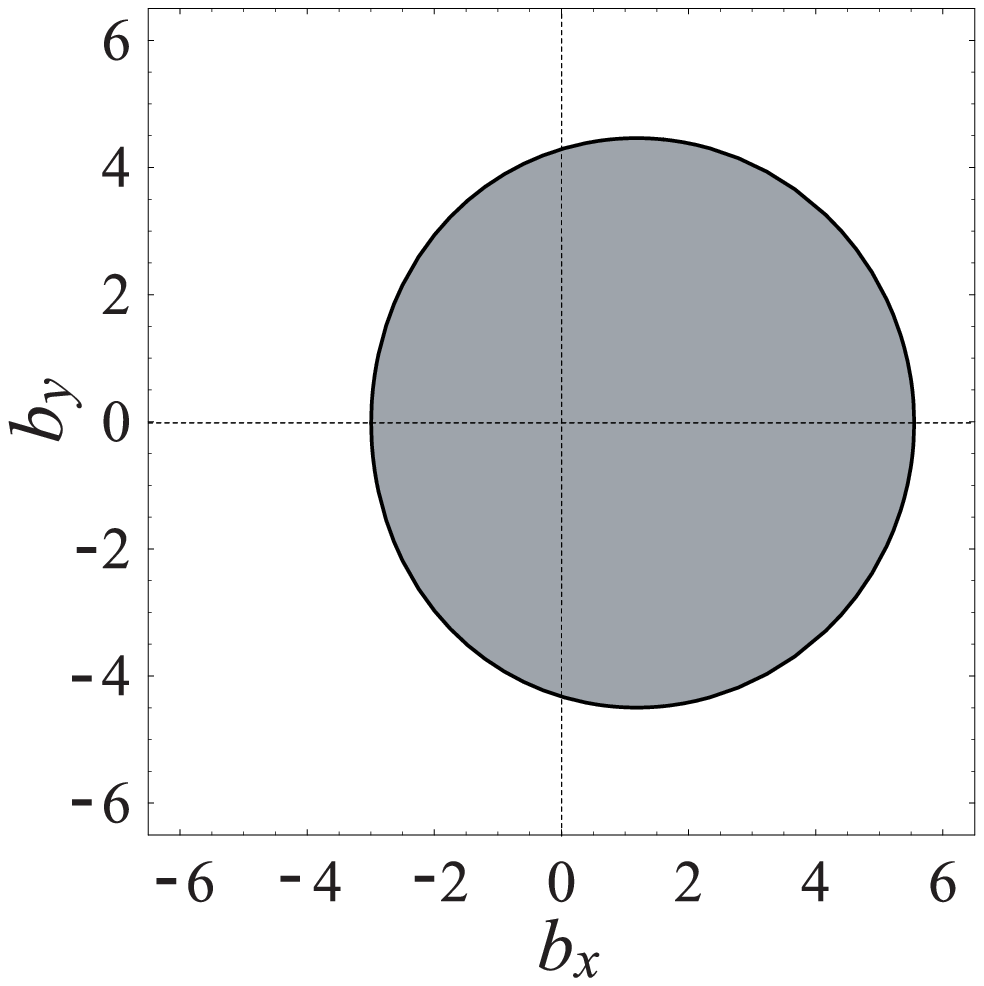} &
			\includegraphics[width=3.5cm]{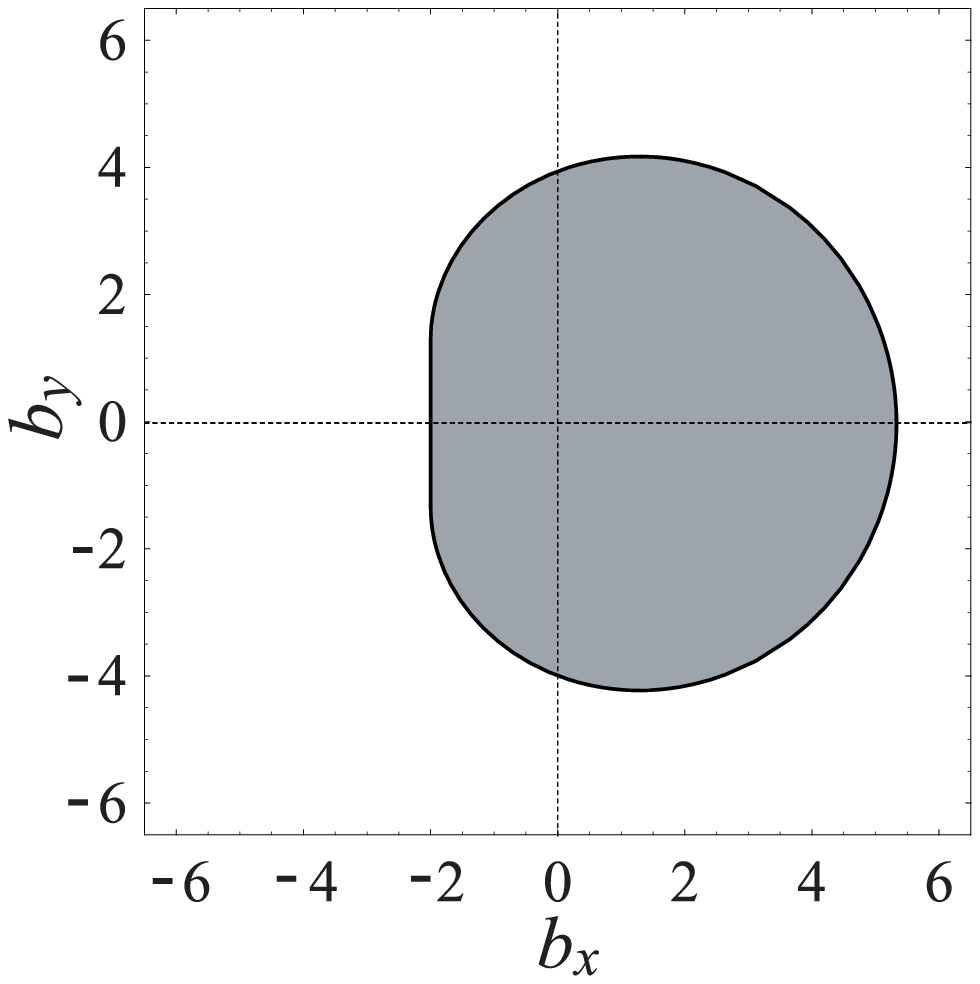} \\
			(a) $Q/M=0$ (BH)&
			(b) $Q/M=\sqrt{3}/2$ (BH)&
			(c) $Q/M=1$ (BH)\\
		\end{tabular}
\caption{\textsf{\footnotesize{Comparison of the shapes of the capture region among various charges for the Sen black hole. The angular momentum and inclination angle are common for all cases, $J/M^2=1/2$ and $i=\pi /2$.  The unit of the coordinates is the mass $M$. One sees that the capture region deviates from a round disk as the charge increases.
} } }				
		\label{fg:fig5}
%----------------------------------------------------------------------%
\vspace{.5cm}
%----------------------------------------------------------------------%
		\setlength{\tabcolsep}{ 70 pt }
		\begin{tabular}{ cc }
			\includegraphics[width=3.5cm]{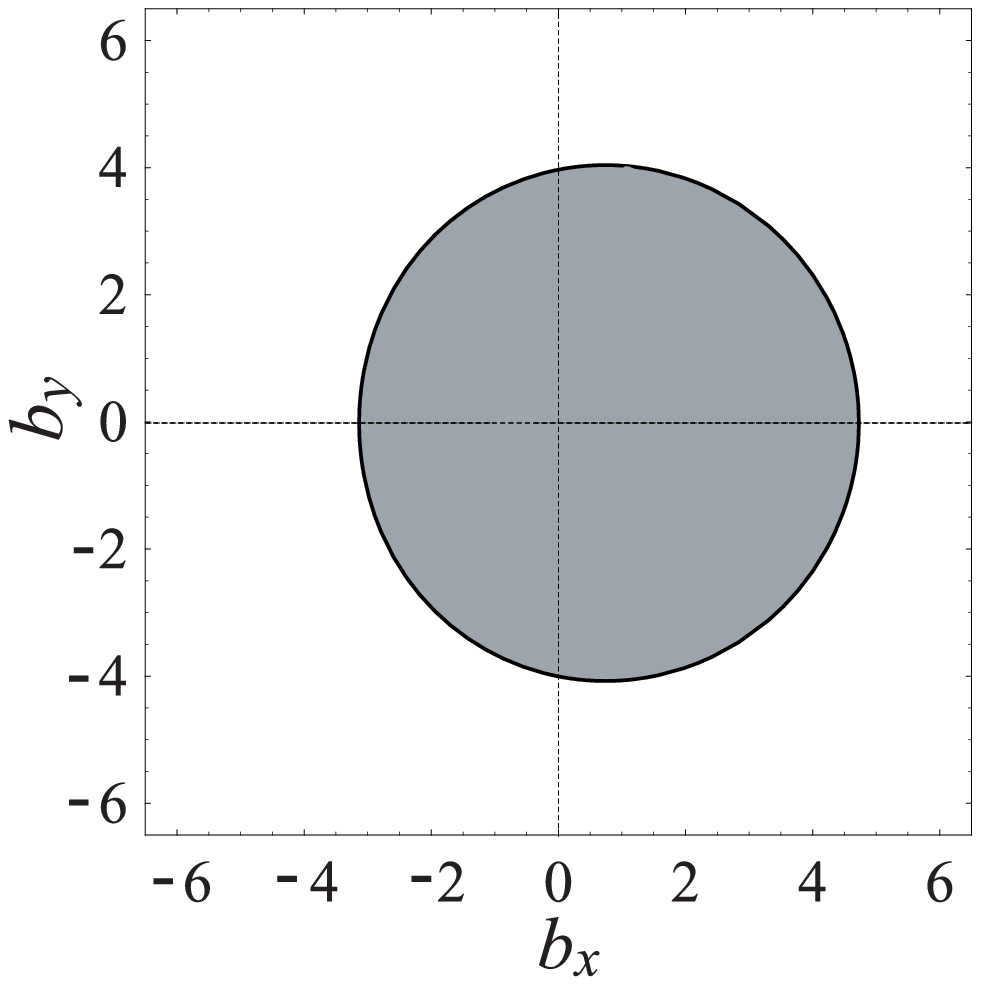} &
			\includegraphics[width=3.5cm]{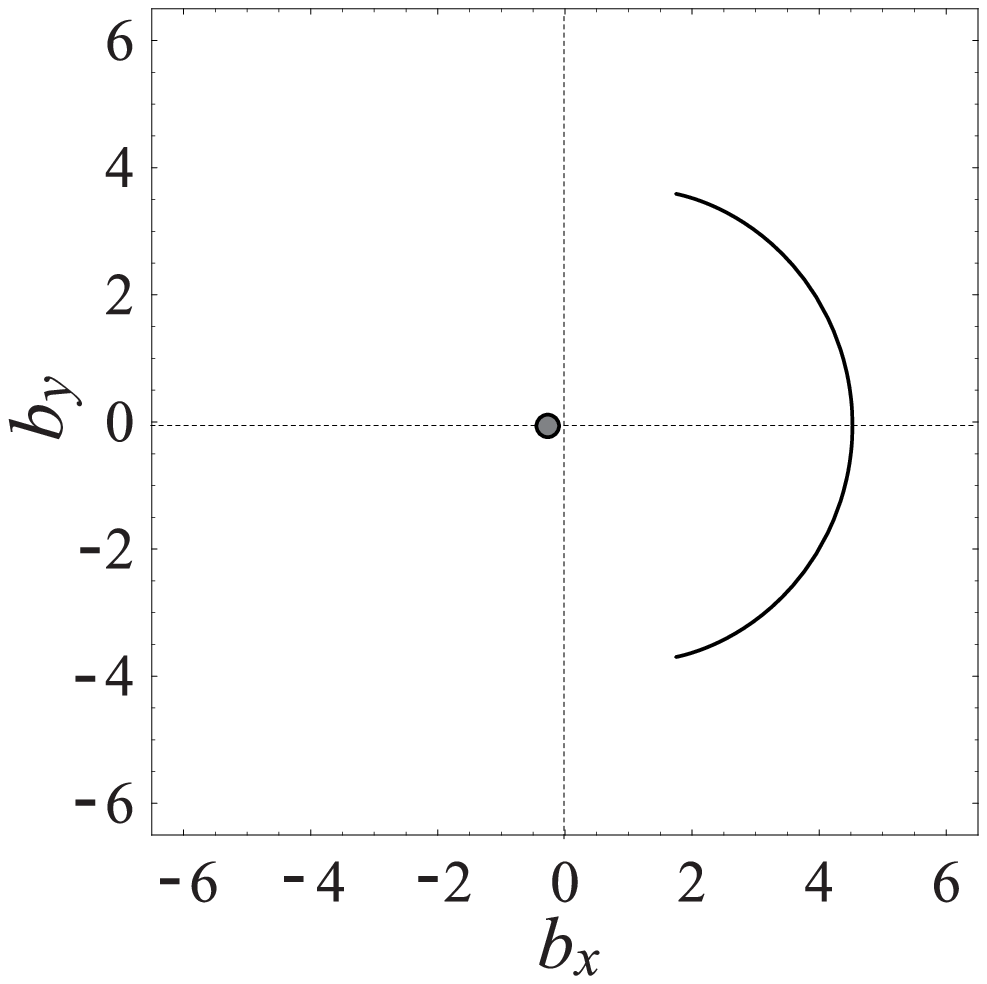} \\
			(a) Sen (BH) &
			(b) KN (NS)\\
		\end{tabular}
\caption{\textsf{\footnotesize{Comparison of the shapes of the capture region between the Sen black hole, (a), and the Kerr-Newman naked singularity, (b), with the common parameters $Q/M =Q_\mathrm{KN}/M_\mathrm{KN} =1$, $J/M^2=J_\mathrm{KN}/M_\mathrm{KN}^2=1/2$, and $i=\pi /6$. The capture region in (a) is a deformed disk. While the capture region in (b) is an arc with a small deformed disk. The arc part is formed by the circular photon orbits located at $r>0$, while the boundary of the small deformed disk is formed by the unstable circular photon orbits located at $r<0$.
	\label{fg:fig6}
} } }				
	\end{figure}
\end{center}
%----------------------------------------------------------------------%
%----------------------------------------------------------------------%

%---------------------------------------------------------------------%
\subsection{Rotating case} 
\label{subsec:rotating}
%---------------------------------------------------------------------%

In the rotating case ($J \not = 0$), the mechanism of the capture and scattering of photons can be understood by the behaviors of potential in a similar manner to the nonrotating case. In this case, however, we have a variety of shapes of shadow in addition to those obtained in the nonrotating case.

Some examples of the capture region in the Sen black holes with rotation are shown in Fig.~\ref{fg:fig5}. Since the conserved quantities of the unstable circular photon orbit, $(\xi_\mathrm{circ},\eta_\mathrm{circ})$, were given in Eq.~(\ref{eq:xietaco}), the set of impact parameters ($b_x,b_y$) for the circular photon orbit can be obtained by substituting Eq.~(\ref{eq:xietaco}) into Eq.~(\ref{eq:imapactparas}). Again, the boundary of shadow is formed by the unstable circular photons orbits, while the inner region of shadow is formed by the photons captured by the event horizon. As a result, the capture region in the Sen black holes with rotation is the deformed disk.

Finally, we illustrate that the Sen and KN solutions having a same extremality can have completely different shapes of shadow as Fig.~\ref{fg:fig6}. In the Sen case the spacetime is a black hole and the shadow is the deformed disk as Fig.~\ref{fg:fig6}(a), while in the KN case the spacetime is a naked singularity and the shadow is an arc with a deformed disk as Fig.~\ref{fg:fig6}(b).
To understand the latter naked singularity case, we should carefully pay attention to the global structure of the spacetime since the maximally extended spacetime has the negative $r$ region where there exists an unstable circular photon orbit. The arc is formed by the unstable circular orbits located at $r>0$~\cite{AdeVries:2000a}, while the boundary of the deformed disk (i.e., the deformed circle) is formed by the unstable circular photon orbits located at $r<0$. Although no event horizon exists, the null rays corresponding to the inner region of the deformed circle cannot go back to the positive $r$ region. Thus, the shadow is the arc with the deformed disk.

%---------------------------------------------------------------------%
%---------------------------------------------------------------------%
\section{Summary} 
\label{sec:conclusion}
%---------------------------------------------------------------------%
%---------------------------------------------------------------------%

We have investigated the null geodesics, the principal null geodesics, and circular photon orbits, in the Sen black hole and also the hidden symmetries which are closely related to the separability.
As the application of the analysis of null geodesics, we have investigated the gravitational capture of photons by the Sen black hole and other charged/rotating black holes and naked singularities in the Kerr-Newman class.

More specifically, we may summarize our results as follows. We obtained the irreducible Killing tensor Eq.~(\ref{eq:Killing}) and conformal Killing tensor Eq.~(\ref{eq:CK}) in the Sen black hole. Then, we showed that the Killing tensors in the Sen black hole have the separable structure by Eqs.~(\ref{eq:SH}), (\ref{eq:Lee}), and (\ref{eq:Kl}), which guarantee the separability of variables. By this observation, it became clear why the separability of variables is possible in the Hamilton-Jacobi/Klein-Gordon equations. We also obtained the principal null geodesics and circular photon orbits. In addition, we found that the principal null vectors, Eq.~(\ref{eq:l}), are the eigenvectors of the covariant derivative of the Killing vector by Eq.~(\ref{eigen}), which plays an important role to analyze the extended (test) objects such as a string. We investigated the capture and scattering of photons by the charged/rotating black holes and naked singularities, described by the Sen and Kerr-Newman solutions, and presented several examples which tell us the characteristic differences of the capture region due to the spacetime structures. The shapes of the capture region were classified into the several types such as the disk, circle, dot, arc, and their combinations. As an example of the difference between the GM and RN solutions, the naked singularity in the GM solution (as the nonrotating limit of the Sen solution) forms the dot in the impact parameter space, Fig.~\ref{fg:fig2}(c), while the naked singularity in the RN solution forms the circle and dot, Fig.~\ref{fg:fig3}(b). It would be interesting to investigate the shadows of naked singularities more in detail from the viewpoint of a connection between astrophysical observations and the cosmic censorship hypothesis~\cite{Virbhadra:2002ju}.

It might be said that the studies of hidden symmetries and separability in four or higher dimensional black holes had been restricted to vacuum solutions. Thus, it is important to generalize such analyses to black holes with charges, as done in this paper. In this paper, we did not consider the naked singular case in the Sen solution since the global structure of the over-extreme Sen solution has not been known. To know the global structures and shadows in this case would help us to know the rich properties of black holes/naked singularities with the rotation and charges.

%---------------------------------------------------------------------%
%---------------------------------------------------------------------%
\section*{Acknowledgments} 
%---------------------------------------------------------------------%
%---------------------------------------------------------------------%
We would like to thank Kei-ichi Maeda for useful discussions.
U.M. was supported by the Golda Meir Fellowship, Israel Science Foundation Grant No.607/05, and DIP Grant No. H.52.

%---------------------------------------------------------------------%
%\bibliographystyle{JHEP}
%\bibliography{GL.bib}
\bibliographystyle{unsrt}
\input{HM-Refs.tex}
%---------------------------------------------------------------------%

\end{document}

%% file: HM-Refs.tex
\providecommand{\href}[2]{#2}\begingroup\raggedright\endgroup